\documentclass[structabstract]{aa}  
\usepackage{graphicx}
\usepackage{natbib}
\usepackage{txfonts}
\usepackage{aalongtable}
\begin{document}
   \title{Peering through the veil: near-infrared photometry and extinction for the
     Galactic nuclear star cluster}

   \subtitle{Accurate near infrared $H$, $Ks$, and $L'$ photometry and
     the  near-infrared extinction-law toward the
     central parsec of the Galaxy}

   \author{R. Sch{\"o}del\inst{1}
          \and
          F. Najarro\inst{2}
          \and
          K. Muzic\inst{3}
          \and
          A. Eckart\inst{3}
          }
   \institute{Instituto de Astrof\'isica de Andaluc\'ia -- CSIC,
     Glorieta de la Astronom\'ia S/N, 18008 Granada, Spain\\
     \email{rainer@iaa.es}
   \and
   Centro de Astrobiolog{\'i}a (CSIC/INTA), Instituto
     Nacional de T\'ecnica Aeroespacial, Ctra de Torrej\'on a Ajalvir,
     km 4, 28850 Torrej\'on de Ardoz, Madrid, Spain\\ \email{najarro@damir.iem.cisc.es}
   \and
   I. Physikalisches Institut, Universit{\"a}t zu K{\"o}ln, Z{\"u}lpicher Strasse 77,
50937 K{\"o}ln, Germany\\  \email{muzic,eckart@ph1.uni-koeln.de}
    }
    
   \date{;;}

 
  \abstract
  {The nuclear star cluster of the Galaxy is an important template for
    understanding its extragalactic counterparts, which can currently not be
    resolved into individual stars. Important drawbacks of
    observations of the Galactic center are, however,  the presence of strong and
    spatially highly variable interstellar extinction and extreme
    crowding of the sources, which makes the use of adaptive optics
    techniques necessary. Both points pose serious obstacles to
    precise photometry that is needed for analyzing the stellar population.}
   {The aims of this work are to provide accurate photometry in multiple
     near-infrared broadband filters, to determine the power-law index
     of the extinction-law toward the central parsec of the Galaxy, to provide
     measurements of the absolute extinction toward the Galactic
     center, and finally to measure the spatial variability of extinction on arcsecond scales.}
   {We use observations of the central parsec of the Milky Way that
     were obtained with the near-infrared camera and adaptive optics
     system NAOS/CONICA at the ESO VLT unit telescope~4. The
     photometric method takes into account anisoplanatic effects and
     limits the corresponding systematic uncertainties to
     $\lesssim2\%$.  Absolute values for the extinction in the $H$,
     $Ks$, and $L'$-bands as well as of the power-law indices of the
     $H$ to $Ks$ and $Ks$ to $L'$ extinction-laws are measured based
   on the well-known  properties of red clump stars. Extinction maps are derived based
     on $H-Ks$ and $Ks-L'$ colors.}
   { We present $Ks$-band photometry for $\sim7700$ stars, and
       additionally photometry for stars detected in the $H$ and/or
       $L'$-bands.  From a number of recently published values we
       compute a  mean distance of the Galactic center of
       $R_{0}=8.03\pm0.15$\,kpc, which has an uncertainty of just
       $2\%$. Based on this $R_{0}$ and on the RC method, we derive
       absolute mean extinction values toward the central parsec of
       the Galaxy of $A_{H}=4.48\pm0.13$\,mag,
       $A_{Ks}=2.54\pm0.12$\,mag, and $A_{L'}=1.27\pm0.18$\,mag. We
       estimate values of the power-law indices of the extinction-law
       of $\alpha_{H-Ks}=2.21\pm0.24$ and
       $\alpha_{Ks-L'}=1.34\pm0.29$.  A $Ks$-band extinction map for
       the Galactic center is computed based on this extinction law and
       on stellar $H-Ks$ colors. Both its statistical and systematic
       uncertainties are estimated to be $<10\%$. Extinction in this
       map derived from stellar color excesses is found to vary on
       arcsecond scales, with a mean value of
       $A_{Ks}=2.74\pm0.30$\,mag. Mean extinction values in a circular
       region with $0.5"$ radius centered on Sagittarius\,A* are
       $A_{H, SgrA*}=4.35\pm0.12$, $A_{Ks, SgrA*}=2.46\pm0.03$, and
       $A_{L', SgrA*}=1.23\pm0.08$. }
   {}

   \keywords{}

   \maketitle
%

\section{Introduction}

Nuclear star clusters (NSCs) are located at the photometric and
dynamical centers of almost all spiral galaxies. Their detailed
  study has only become possible via {\it HST} observations in the
1990s, but they are of great interest because they are the densest and most
massive clusters in their host galaxies. Frequently they contain
supermassive black holes at their centers.  They show signs of
recurring star formation, and the discovery of rotation of the NSCs of
NGC\,4244 \citep{Seth:2008kx} and of the Milky Way
\citep{Trippe:2008it,Schodel:2009zr} indicates that NSCs may at least
partially form by accretion of material from their host galaxy. A
concise review of the properties of NSCs and hypotheses on their
formation is given in \citet{Boker:2008nx}.

Studying the NSC at the Galactic center (GC) is of great importance
for the understanding of NSCs in general because in the next decades
it will remain impossible to resolve any but the brightest individual
stars in extragalactic NSCs, even with 50m-class telescopes. A further
reason for the importance of the GC NSC is that the existence of the
supermassive black hole, Sagittarius\,A*, at the GC is well
established and the mass of this black hole has been determined with
high precision \citep[e.g.,][]{Ghez:2008oq,Gillessen:2009qe}. Strong
interest in the stellar population at the GC was raised by
spectroscopic observations, which have revealed the existence of
numerous massive, young stars within 0.5\,pc of Sagittarius A*. Those
young stars are witnesses of a starburst event that occurred about 4
million years ago. The young stars are arranged in a non-isotropic
pattern and at least 50\% of them are located in a rotating disk
\citep[see][]{Bartko:2009fq,Lu:2009bl}.

Unfortunately, there are two important drawbacks in observational
studies of the Milky Way's NSC: extreme crowding and high
extinction. While the great progress made in adaptive optics
techniques at 8-10m-class telescopes in the past years has been highly
successful in reducing the problem of crowding, it nevertheless
remains a serious issue. Even with instruments such as NACO at the ESO
VLT, strong crowding limits the point-source sensitivity within
$\sim0.5$\,pc of Sagittarius\,A* so significantly that only stars more
massive than $\sim2\,M_{\odot}$ can be detected in this region. This
means that we can still detect only of the order $10\%$ of the stars
in the NSC \citep{Schodel:2007tw}. Additionally, it is notoriously
difficult to extract accurate photometric measurements from AO
observations of crowded fields, particularly when the field-of-view
exceeds the size of the isoplanatic angle.

As concerns extinction, we can consider ourselves lucky that we can
observe the GC at all. Within 2-3\,pc of Sagittarius\,A* there exist
plenty of clouds that are almost or completely opaque even in the
near-infrared \citep[for example, see Fig.\,1
in][]{Schodel:2007tw}. Average extinction toward the central parsec in
the K-band reaches $A_{K}\approx3$. As concerns extinction at visual
wavelengths, \citet{Rieke:1985fq} found $A_{V}/A_{K}\sim9$, but more
recent studies indicate higher values of $A_{V}/A_{K}\sim16$
\citep{Nishiyama:2008qa}, or even $A_{V}/A_{K}\sim29$
\citep{Gosling:2009kl}. Therefore, the stellar population at the GC
becomes observable only at near-infrared wavelengths. Only the
brightest stars can be seen in the $J$-band. Reasonable S/N on the
stars is only reached in the $H$, $K$, and $L$ bands. This restriction
to a tight wavelength window, where most stars show only small color
indices, combined with the fact that the extinction toward the GC is
not only high, but also variable on small spatial scales of a few
arcseconds \citep{Scoville:2003la,Gosling:2006eq,Schodel:2007tw}, makes photometric
studies of the stellar population at the GC extremely difficult.

Although adaptive optics assisted integral field spectroscopy has
resulted in the classification of hundreds of stars within
$\sim0.5$\,pc of Sagittarius\,A*
\citep[e.g.,][]{Maness:2007sj,Paumard:2006xd,Do:2009tg}, this technique is extremely time
consuming in both observations and data analysis. In order to
efficiently probe large fields in the Galactic center for the presence
of young, massive stars, it is therefore desirable to improve the
photometric observations in a way that enables us to at least crudely
distinguish whether a star in the GC is of early or of late spectral
type. How this can be done in spite of the difficulties described
above has been demonstrated recently by \citet{Buchholz:2009sp}. They
used intermediate band filters across the K-band, calibrated
photometry locally with the help of red clump stars,  and used the CO
bandhead absorption feature of late-type stars as a distinguishing
criterion.

Accurate knowledge of extinction has impact in many different aspects
of GC research: the stellar surface number density is not only a
function of distance from Sagittarius\,A*, but is also 
dependent on local extinction \citep{Schodel:2007tw}; extinction is an
important ingredient for modeling stellar atmospheres
\citep[e.g.,][]{Najarro:1997qe,Martins:2007sf}; extinction in the
infrared is also important for estimating the intrinsic brightness of
the near-infrared emission from the accretion flow or outflow related
to Sagittarius\,A*
\citep[e.g.,][]{Genzel:2003hc,Eckart:2006sp,Do:2009ij}. The mentioned
examples show that one must know not only the absolute value of
extinction, but also how it changes as a function of wavelength and
how it varies on spatial scales as small as one to a few arcseconds.

Until just about a decade ago, the extinction law from near- to
mid-infrared wavelengths ($\sim1-8\,\mu$m) was thought to be described
well by a ``universal'' power-law, $A_{\lambda}\propto
\lambda^{-\alpha}$, with a power-law index of $\alpha\approx1.75$
\citep[e.g.,][]{Draine:1989eq}. First doubts were shed on this model
by spectroscopic observations of the GC with ISO SWS that showed clear
deviations from this model at wavelengths $\gtrsim3\,\mu$m
\citep{Lutz:1996oz}. Subsequent studies have confirmed that the near-
to mid-infrared extinction-law is more complex than a simple
power-law. While it seems well established by now that a simple
power-law is an adequate description of the wavelength dependence of
extinction in the near-infrared, albeit with a stronger wavelength
dependence as previously thought, the situation is less clear longward
of $\sim3\,\mu$m. Recent investigation shows that the extinction law
decreases much less rapidly at $\lambda>3\,\mu$m than suggestd by
earlier estimates and may actually be almost constant at $5-10\,\mu$m
\citep[see][for an extensive discussion of these
points]{Nishiyama:2009oj}. Taking the problem even a step further, in
a recent study of extinction toward the nuclear bulge
\citet{Gosling:2009kl} find that the index of the power-law
extinction-law can change on spatial scales as small as
5\,arcseconds. This signifies that we may be forced to abandon the
concept of a universal extinction-law.

It is evident that the topic of extinction toward the central parsec
of the Galaxy needs to be re-visited. Extinction maps for the central
0.5-1\,pc were recently presented by \citet{Schodel:2007tw} and
\citet{Buchholz:2009sp}.  However, these papers were not mainly
focused on extinction and the extinction maps were rather
side-products. Both papers used the old $A_{\lambda}\propto
\lambda^{-1.75}$ extinction-law \citep{Draine:1989eq}. Neither the
extinction law nor  absolute extinction were examined in depth. Also,
the absolute photometric calibration in our previous works was of limited
accuracy because the PSF for the stars was either truncated
\citep{Schodel:2007tw} or because complex local calibration procedures
had to be adopted \citep{Buchholz:2009sp}.

The aim of this paper is to carry us a step forward in observational
studies of the GC NSC by addressing two of the above described
problems: accurate photometry with AO over a large FOV and accurate
measurements of extinction.  Schoedel (A\&A, 2009, accepted for
publication) has recently presented a method for the photometric
analysis of AO data with spatially variable PSF, but a very limited
number of suitable PSF reference stars. Based on this method, in this
paper we analyze imaging observations of the GC in the $H$, $Ks$, and
$L'$ bands and provide accurate photometry (with both statistical and
systematic uncertainties of the order $\sim5\%$ at $Ks$) for several
thousand sources observed in these bands within $\sim$1\,pc of
Sagittarius\,A*. We determine both the absolute extinction and the
extinction law at the various bands and provide a detailed extinction
map (resolution of $1"-2"$) for the central parsec of the GC.

\begin{figure}[!htb]
\includegraphics[width=\columnwidth,angle=0]{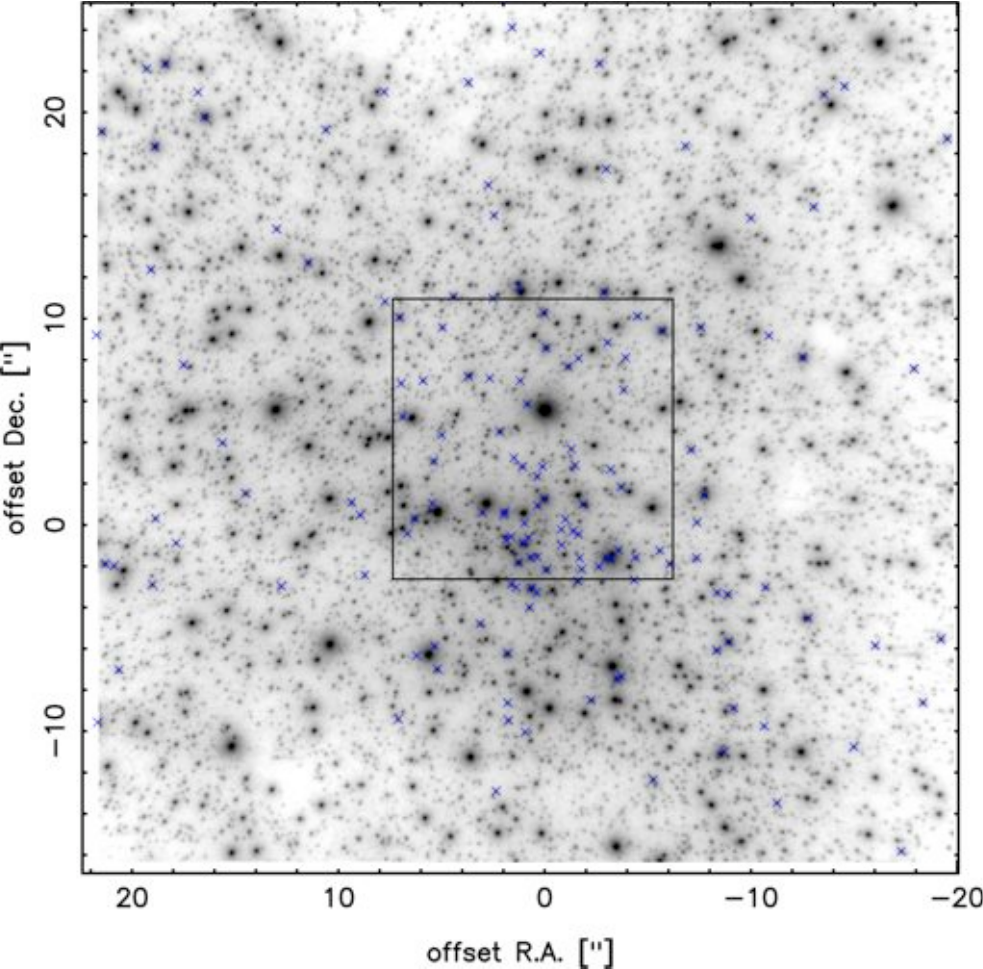}
\caption{\label{Fig:mosaicKs} Mosaic image of the $Ks$-band
  observations. North is up and east is to the left. Offsets are given
  in arcseconds from Sgr\,A*. The guide star, IRS\,7, is the brightest
  star within the rectangle, about $5.5"$ north of Sgr\,A*. The black
  rectangle indicates the area of overlap between the 4 dither
  positions. Blue crosses mark early-type stars identified by their
  $H-Ks$ colors with extinction/reddening correction applied (see
  section\,\ref{sec:extmap} and Fig.\,\ref{Fig:hkcorr}).}
\end{figure}



\section{Observations}

\begin{figure}[!htb]
\includegraphics[width=\columnwidth,angle=0]{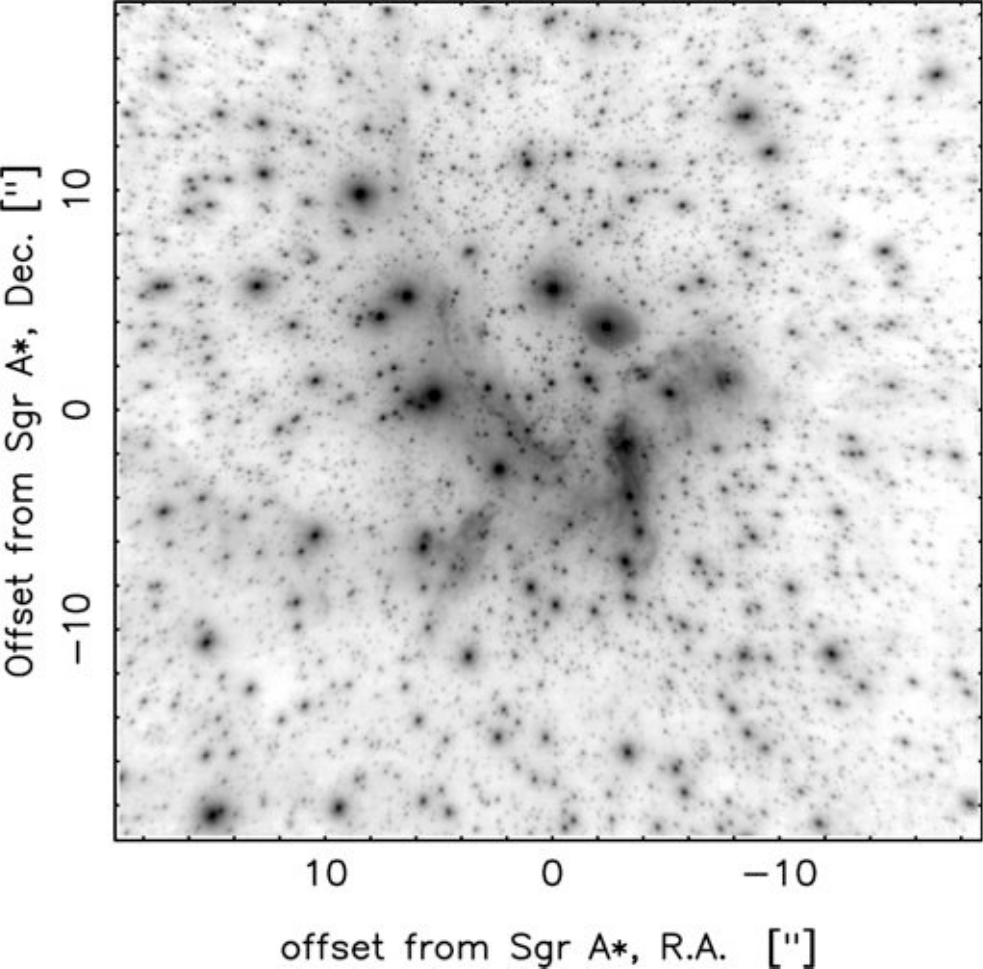}
\caption{\label{Fig:mosaicLp} Mosaic image of the $L'$-band
  observations. North is up and east is to the left. Offsets are given
  in arcseconds from Sgr\,A*.}
\end{figure}

The observations used in this work were obtained with the
near-infrared camera and adaptive optics (AO) system NAOS/CONICA
(short NACO) at the ESO VLT unit telescope~4 \footnote{Based on
  observations made with ESO Telescopes at the La Silla or Paranal
  Observatories under programmes ID 077.B-0014 and 081.B-0648}. The $\rm
Ks\approx6.5-7.0$ supergiant IRS\,7 was used to close the loop
of the AO, using the NIR wavefront sensor. The sky background was
measured on a largely empty patch of sky, a dark cloud about $400''$
north and $713''$ east of the target. Sky subtraction, bad pixel
correction, and flat fielding were applied to the individual
exposures. The NACO S27 camera, with a pixel scale of $0.02705''/{\rm
  pix}$, was used for both $H$ and $Ks$ band observations. The
field-of-view (FOV) of a single exposure is thus $28''\times28''$. The
observations were dithered by applying a rectangular dither pattern
with the center of the dithered exposures positioned approximately at
$(8.0'',-2.6'')$, $(-6.1'',-2.7'')$, $(-6.1'',11.2'')$, and
$(8.1'',11.3'')$ east and north of Sgr\,A*. In the text we refer to
these four offsets as dither positions 1, 2, 3, and 4. The combined
FOV of the observations is about $40''\times40''$ and is offset to the
north with respect to Sgr\,A* because the guide star IRS\,7 is located
about $5.6''$ north of Sgr\,A*.

Seeing in the optical ranged between $0.6"$ and $1.0"$ for both $H$ and $Ks$
observations.  According to the ESO Ambient Conditions Database
  \footnote{\it http://archive.eso.org/asm/ambient-server?site=paranal} sky
  transparency variations during the observations were below $1\%$.
The achieved Strehl ratio ranged between $\sim20\%$ ($\sim15\%$) near
the guide star and $\sim8\%$ ($\sim8\%$) at $25"$ distance from the
guide star in the $Ks$-($H$-)band. The Strehl ratio was estimated
using the {\it Strehl} algorithm of the ESO {\it eclipse} software
package \citep{Devillard:1997kx} on PSFs extracted at various
positions in the image. From the multiple measurements we estimate the
$1\,\sigma$ uncertainty of the measured Strehl ratio to $\sim3\%$.
Table\,\ref{Tab:Obs} summarizes the observations.  The detector
integration time (DIT) was set to $2.0$\,s in order to avoid
saturation of the brightest stars. The central pixels of a small
number of the brightest stars were saturated in $Ks$. Their cores were
repaired using nearby unsaturated stars and the PSF extraction routine
of {\it StarFinder}. After 28\,DITs, the instrument averaged the data
to a single exposure (NDIT$=28$). In this way, 8 individual exposures
were obtained per dither position. The exposures of each respective
dither position were aligned (to compensate for small residual shifts)
with the {\it jitter} algorithm of the ESO {\it eclipse} software
package \citep{Devillard:1997kx}. We show the combined FOV of the
$Ks$-band observations in Fig.\,\ref{Fig:mosaicKs}. Note that the
photometry and astrometry in this work was done on the combined images
of {\it each dither position} and {\it not on the combined mosaic} of
all images (as shown in Fig.\,\ref{Fig:mosaicKs}) in order to have a
constant signal-to-noise ratio over the entire images. The
$\sim13.5"\times13.5"$ overlap area between the four dither positions
is indicated by the rectangle in Fig.\,\ref{Fig:mosaicKs}.

Seeing in the optical ranged between $1.0"$ and $2.0"$ for the $L'$
observations.  The NACO L27 camera, with a pixel scale of
$0.02719''/{\rm pix}$, was used for the $L'$ observations. According
to the ESO Ambient Conditions Database sky transparency variations
during the observations were below $1\%$.  The observations were
random jittered, with a jitter box width of $20"$. Offsets by $60"$ in
random directions were alternated with observations of the target. The
frames taken on the offsets frames, where the stellar density is
lower, were used to construct sky frames. The 7 offset frames nearest
in time to a given exposure on the target were median combined and
subtracted as sky measurement. Flat fielding and bad pixel correction
were applied.  Finally, all exposures were combined with the {\it
  jitter} algorithm of the ESO {\it eclipse} software package. The
resulting mosaic image is shown in Fig.\,\ref{Fig:mosaicLp}.

\begin{table}
\centering
\caption{Details of the imaging observations used in this
  work.}
\label{Tab:Obs} 
\begin{tabular}{llllll}
\hline
\hline
Date & $\lambda_{\rm central}$ & $\Delta\lambda$ & N$^{\mathrm{a}}$ & NDIT$^{\mathrm{b}}$ & DIT$^{\mathrm{c}}$\\
 &  [$\mu$m]  &   [$\mu$m] &  & & [s] \\
\hline
29 April 2006 & 1.66 & 0.33 & 32 & 28 & 2  \\
29 April 2006 & 2.18 & 0.35 & 32 & 28 & 2 \\
26 May 2008 & 3.80 & 0.62 & 202 & 150 & 0.2\\
\hline
\end{tabular}
\begin{list}{}{}
\item[$^{\mathrm{a}}$] Number of (dithered) exposures
\item[$^{\mathrm{b}}$] Number of integrations that were averaged on-line by the read-out
  electronics
\item[$^{\mathrm{c}}$] Detector integration time. The total integration time of each observation amounts to N$\times$NDIT$\times$DIT.
\end{list}
 \end{table}


\section{Photometry}

\begin{figure}[!tbh]
\includegraphics[width=.9\columnwidth,angle=0]{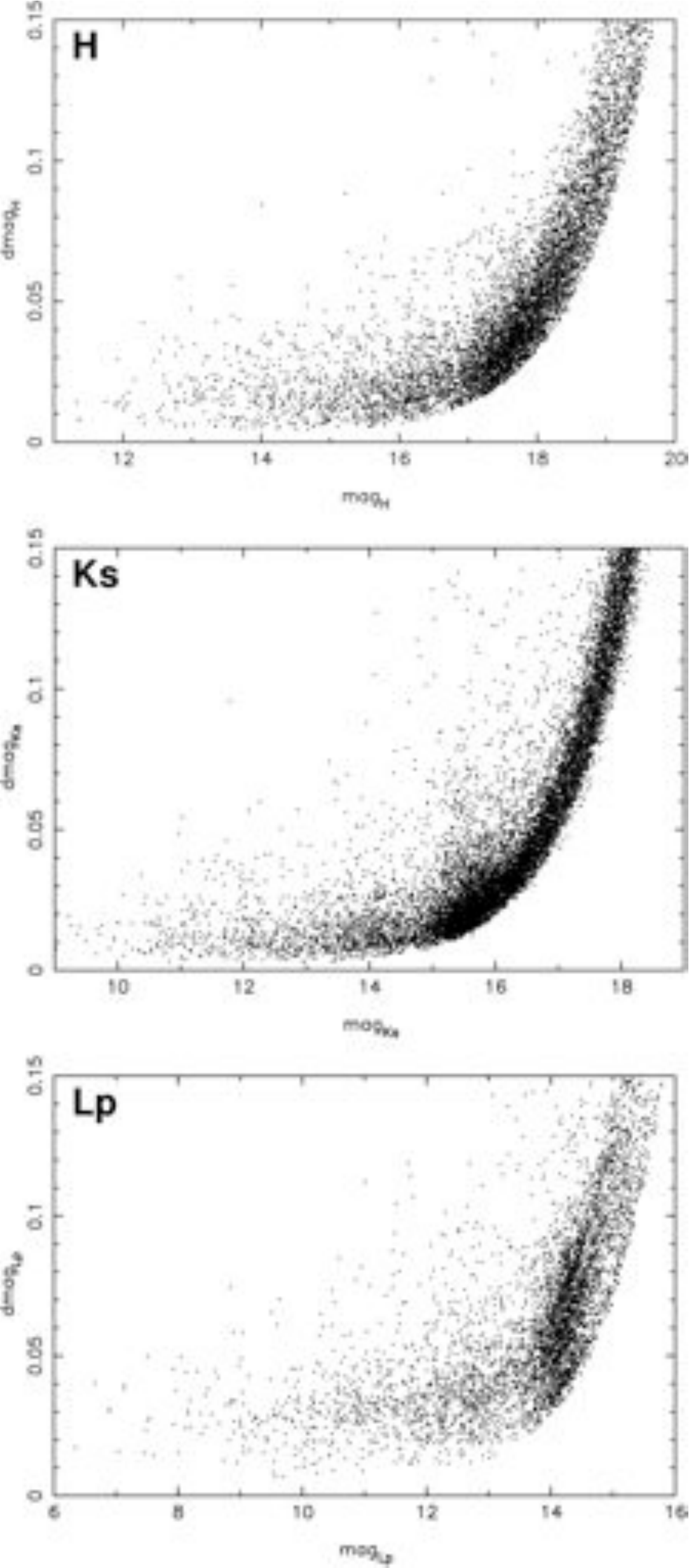}
\caption{\label{Fig:hklphot} Statistical photometric uncertainty vs.\ magnitude
  for the $H$- (top), $Ks$- (middle), and $L'$- (bottom) bands.}
\end{figure}

Due to the extreme source density at the GC aperture photometry cannot
be applied accurately to GC imaging data. Therefore PSF fitting
techniques must be used. In AO imaging anisoplanatic effects can lead
to a systematic bias in the photometry of point sources when PSF
fitting is used for estimating the flux of the stars. This bias depends
on various factors such as distance from the guide star, observing
wavelength, and atmospheric conditions at the time of the
observations. In case of the $H$-band data used in this work the
systematic error of the photometry can reach $0.25$\,mag across the
FOV, if a single, non-variable PSF is used (see Schoedel, A\&A,
  2009, accepted for publication). For example, anisoplanatic effects
make stars at distances larger than the isoplanatic angle appear
elongated, with their major axes pointing toward the guide star. This
can be nicely seen in the image shown in Fig.\,\ref{Fig:mosaicLp}
(note: usually anisoplanatic effects are stronger at shorter
wavelengths, but in the observations used for this paper, it happens
that those effects are stronger in the $L'$-band than in the $Ks$-band
image). Various methods for dealing with anisoplanatic effects have
been suggested. A straightforward and effective method to deal with
anisoplanatic effects is presented in a companion paper (Schoedel,
A\&A, 2009, accepted for publication).  It not only provides accurate
point source photometry but also reliable estimation of the diffuse
background emission. In brief, the method consists of two steps: (a)
Wiener deconvolution, using a point spread function (PSF) extracted
from the guide star (or bright stars near the guide star), in this
case IRS\,7; (b) photometry and astrometry on the deconvolved image
via PSF fitting with the {\it StarFinder} \citep{Diolaiti:2000rz}
program package. The image is partitioned into overlapping sub-frames
that are smaller than the isoplanatic angle. Those sub-frames are
analyzed independently. As shown in Schoedel (A\&A, 2009, accepted for
publication) this
method minimizes the systematic uncertainty of the point source
photometry due to anisoplanatic effects to $\lesssim0.02$\,mag across
the $28"\times28"$ FOV of the images at each dither position.

After deconvolution with PSFs extracted from IRS\,7, the $H$- and
$Ks$-band images from the four dither positions were partitioned into
$13\times13$ overlapping sub-frames of $256\times256$\,pixel$^{2}$
($\sim6.9"\times6.9"$). The shifts between sub-frames were thus just
64\,pixels in x- and/or y-direction to assure large overlap between
the sub-frames.  We used the following {\it StarFinder} parameters:
$min\_corr=0.9$, $thresh=[5.,5.]$, and $back\_box=30$. In order to
avoid spurious detections, point source detection was also run on
non-deconvolved images and only sources that were discovered in both
the deconvolved and non-deconvolved images were included in the final
list of detected stars. The images for the diffuse background and
residuals were obtained by recombining the corresponding
sub-frames. The method provides two kinds of uncertainties for the
photometric and astrometric measurements: {\it formal} uncertainties,
computed by the {\it StarFinder} algorithm from Gaussian and photon
noise, and {\it PSF uncertainties}, resulting from incomplete
knowledge of the local PSF. The latter is calculated by comparing
multiple measurements of the stars because they are present in several
sub-frames. For stars without multiple measurements we adopted a PSF
uncertainty of $0.02$\,mag for the $H$-band and $0.015$\,mag for the
$Ks$-band (see Fig.\,\ref{Fig:hklphot}). The results of multiple
measurements were averaged.  Since deconvolution changes the noise
properties of the image, the formal uncertainties computed by {\it
  StarFinder} are too small (for details see Schoedel,A\&A, 2009, accepted for
publication). Monte Carlo simulations were used to determine a correction
factor with which we scaled the formal uncertainties.  The final
uncertainties were calculated by quadratically combining the re-scaled
formal fit uncertainties with the PSF uncertainties. The data in the
overlap region between the four dither positions were averaged. Stars
present in the overlap region between the  dithered fields in the  $H$
and $Ks$ observations were used to check the photometric
accuracy. Photometry on the stars in the overlapping fields shows that
the  uncertainties estimated by the applied  method are correct (see
Schoedel, A\&A, 2009, accepted for publication).

\begin{figure}[!htb]
\includegraphics[width=\columnwidth]{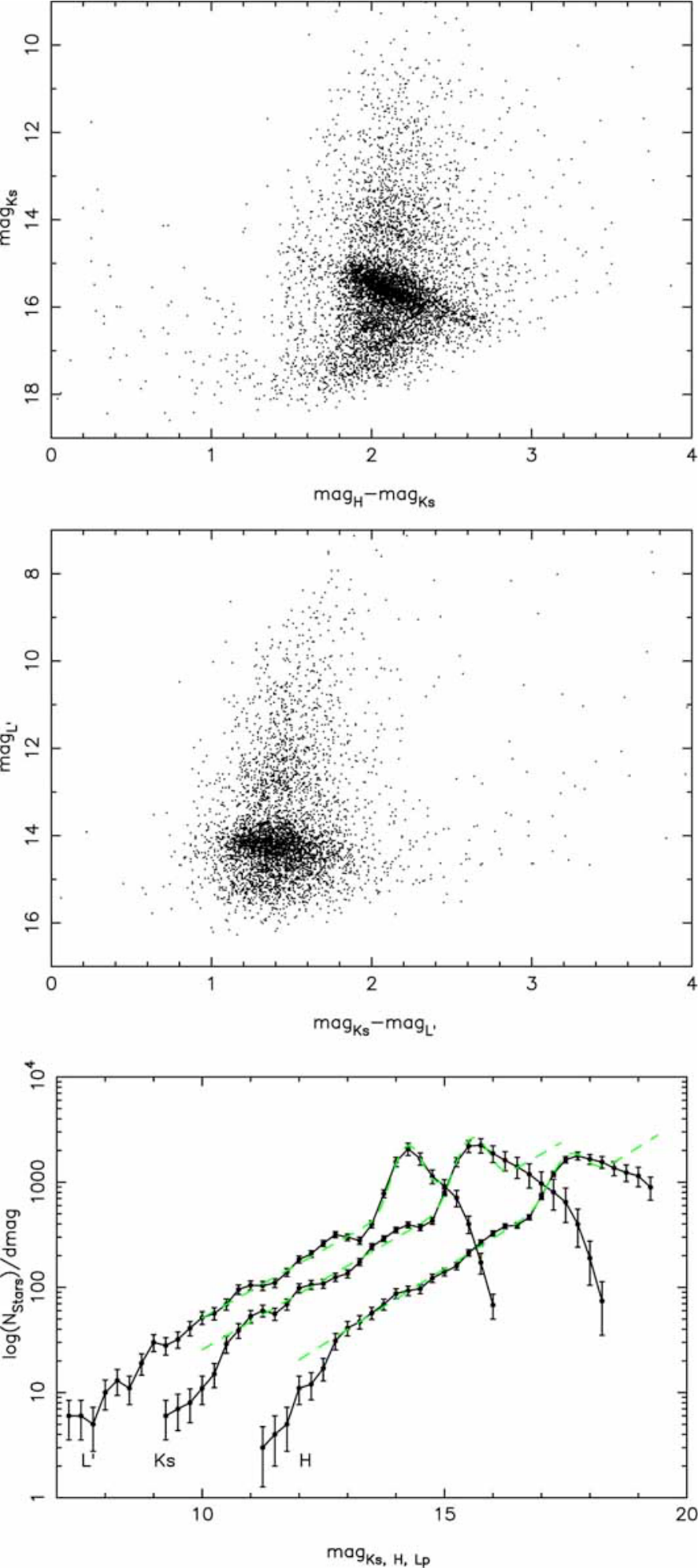}
\caption{\label{Fig:colmag} Color-magnitude diagrams, with $Ks$
  vs.\ $H-Ks$ shown in the top panel and $L'$ vs.\ $Ks-L'$ in the
  middle one. Bottom: Completeness (due to crowding) corrected $L'$, $Ks$, and $H$
  luminosity functions. The dashed green lines are fits with an
  exponential plus a Gaussian.}
\end{figure}

Photometry on the L' image was performed in a very similar way, except
that there was only one combined frame to be examined. The PSF for
initial linear deconvolution was extracted from IRS\,7. The size of
the sub-frames for local PSF fitting was $300\times300$\,pixel$^{2}$,
with a shift of 50 pixels between the frames. Since the $L'$ data are
very rich in structure as concerns the diffuse emission, we set the
$back\_box$ to a smaller value (20\,pixels).

Zero points for the NACO instrument for various combinations of
cameras, filters, and dichroics (that split the light between
wavefront sensor and camera) are determined routinely within the ESO
instrument calibration plan. Zero points for all filters used in this
work and for the corresponding setup (camera S27, dichroic N20C80)
were determined via observation of a standard star during the same
night as the observations: $ZP_{H}=23.64\pm0.05$,
$ZP_{Ks}=22.85\pm0.05$, and $ZP_{L'}=22.38\pm0.15$. Additional
uncertainty of the zero point is caused by the uncertainty of the PSF
that is used for deconvolution. From experiments (choosing different
PSF sizes) we estimate this additional uncertainty to
$\lesssim0.03$\,mag. We combine this additional uncertainty
quadratically with the previous uncertainties and obtain final zero
point uncertainties of $\Delta ZP_{H}=0.06$, $\Delta ZP_{Ks}=0.06$,
and $\Delta ZP_{L'}=0.15$. The annual mean and standard deviation
  of the zero points corresponding to the year and the setups used for
  the observations in this work are $ZP_{H,mean, 2006}=23.61\pm0.10$,
  $ZP_{Ks,mean,2006}=22.82\pm0.07$, and
  $ZP_{L',mean,2008}=22.25\pm0.17$. This agrees well with the zero
  points and corresponding $1\,\sigma$ used for the photometry in this
work.

Plots of photometric uncertainty vs.\ magnitude for all point sources
discovered in the individual bands are shown in
Fig.\,\ref{Fig:hklphot}.  We combined the photometric results to a
final list, requiring that each star must be detected in at least two
different bands.  The final lists of detected point sources was
aligned with the coordinate system of the $Ks$-observations. The full
list of all stars and their fluxes in the different bands is given in
Table\,\ref{Tab:list}, available online.

Because of the non-standard photometric methodology applied here
(deconvolution and extraction of local PSFs) it was not
straightforward to determine incompleteness due to crowding via the
standard method of inserting and recovering artifical stars. The main
difficulty here is the change of the PSF across the FOV.  We therefore
estimated completeness directly from the data using the methodology
suggested by \citet{Eisenhauer:1998tg} and further described in
\citet{Harayama:2008ph}. In brief, it consists of deriving a critical
distance for each possible magnitude difference between two given
stars. A star closer to a brighter one than the critical distance will
not be detected. Once the critical distance is known for all relevant
magnitude differences, completeness maps can be
created. Table\,\ref{Tab:completeness} gives the completeness of our
data for different magnitudes and bands. Some of the assumptions of
the methodology suggested by \citet{Eisenhauer:1998tg} may be violated
in our case, particularly becase the stellar density in the GC FOV
changes strongly because of effects of extinction and cluster
structure. For this reason we estimated the critical distances in
various ways, e.g. by analysing different parts of the FOV. In
Table\,\ref{Tab:completeness} we therefore give two values for
estimated completeness, a high and a low one, for each magnitude and
band. Fortunately it turns out (see following section) that
completeness correction does {\it not} have any significant influence
on our results.

\begin{table}
\centering
\caption{Completeness levels in percent  for given
    bands and magnitudes.}
\label{Tab:completeness} 
\begin{tabular}{lrrr}
\hline
\hline
mag & $H$ & $Ks$ & $L'$ \\
\hline
12 & 100-100$^{\mathrm{a}}$ & 98-100$^{\mathrm{a}}$ & 93-96$^{\mathrm{a}}$ \\
13 & 100-100 & 96-98 & 84-91 \\
14 & 99-99     & 91-97 &  64-77 \\
15 & 97-99     & 83-92 & 54-69 \\
16 & 93-97  & 59-82 & 42 - 58 \\
17 & 83-91  &  47 - 70 & n.a. \\
18 & 66-82  &  28 - 50 & n.a.\\
19 & 50-70 & n.a.  & n.a.\\    
\hline
\end{tabular}
\begin{list}{}{}
\item[$^{\mathrm{a}}$] A high and a low value of estimated
    completeness are given for each combination of band and magnitude.
\end{list}
 \end{table}

\section{Constraining extinction with the red clump}

Color-magnitude diagrams of our data are shown in Fig.\,\ref{Fig:colmag}. The
densely populated area around $Ks\approx15.6$ and
$L'\approx14.2$ is due to red clump (RC) stars, which are the
equivalent to horizontal branch stars in metal-rich populations. The
luminosities and colors of RC stars are narrowly distributed and are
only weakly dependent on age and metallicity, particularly for
clusters older than $\sim$1.6\,Gyr
\citep[e.g.,][]{Castellani:1992kk,Alves:2000lr,Grocholski:2002wq}.

The completeness corrected luminosity functions (LFs) in the $H$,
$Ks$, and $L'$ bands are shown in the bottom panel of
Fig.\,\ref{Fig:colmag}. The RC feature is clearly visible as a bump in
all LFs. We fitted all LFs with the combination of exponentials plus
Gaussians. The locations of the peaks of the RC as derived from these
fits are $H_{\rm RC}=17.60\pm0.04\, (17.58\pm0.016)$, $Ks_{\rm
  RC}=15.59\pm0.04\,(15.57\pm0.009)$, and $L'_{\rm
  RC}=14.25\pm0.03\,(14.22\pm0.008)$. Values in brackets are the
results obtained if no crowding completeness correction is
applied. The uncertainties of the fits are smaller than the
uncertainty of the absolute photometric calibration ($0.06$\,mag for
$H$ and $Ks$, and $0.15$\,mag for $L'$). In the following analysis we
will use the numbers based on the completeness corrected LFs. We note
however, that this crowding completeness corection has a negligible
effect, of the order 10\% of the uncertainty, on the final results. We
attribute this to two reasons: (a) the RC is located in a magnitude
range where completeness is fairly high; (b) the dominating
uncertainty is the one of the absolute photometric calibration. That
incompleteness due to crowding is not a serious issue can also be seen
when comparing the $Ks$LF from this paper with the one shown in
Fig.\,10 of \citet{Schodel:2007tw}. A different, slightly deeper data
set was used for the latter work, but we can see that the RC bump in
the $Ks$LF is negligibly affected by incompleteness. Note, however,
that absolute calibration was not a goal of \citet{Schodel:2007tw},
therefore, there RC peak is shifted by $\sim0.3\,$mag to brighter
magnitudes. Finally, we may ask whether the strong extinction toward
the GC may affect the RC peak in the LFs of the different filters and
thus bias the peak toward brighter values. RC stars are, however, well
within the sensitivity of the observing setup used, at all bands. For
example, an {\it isolated} $Ks=20$ source would be detected with NACO
and the settings used for this work at a level of
$\sim10\,\sigma$. The only incompleteness that affects us here is the
one due to crowding, which has fortunately a negligble effect on our
results.

\begin{figure}[!htb]
\includegraphics[width=\columnwidth]{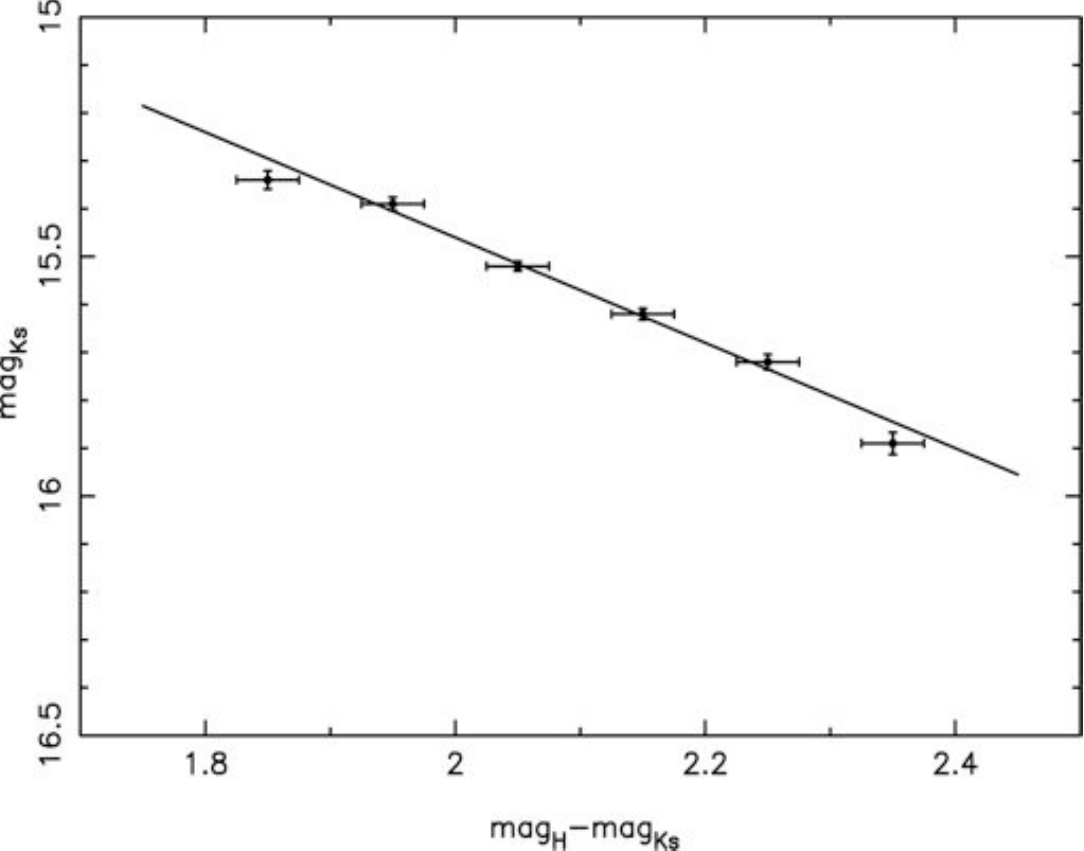}
\caption{\label{Fig:RCfit} Peak $Ks$-magnitude 
  vs.\ $H-Ks$ color of the red clump.}
\end{figure}

Since RC stars can be found in large numbers toward the Galactic bulge
and center, they are highly useful for studying Galactic extinction
and structure
\citep[e.g.,][]{Wozniak:1996kl,Udalski:2003dp,Sumi:2004tg,Nishiyama:2005ao,Nishiyama:2006tx,Nishiyama:2006ai}.
The absolute magnitude of RC stars has been estimated to
$M_{K}=-1.54\pm0.04$ (in the 2MASS system) by
\citet{Groenewegen:2008wj} and to $M_{K}=-1.61\pm0.03$ by
\citet{Alves:2000lr}. Both values are based on the {\it Hipparcos}
catalogue. Here we adopt the more recent value for $M_{K}$.  The
difference between the $K$ and $Ks$-magnitude is $\approx0.01$\,mag
\citep[see discussion in][]{Nishiyama:2006ai}.
\citet{Nishiyama:2006ai} estimated the distance to the GC based on
photometry of RC stars, calculating the distance modulus as
\begin{equation}
(m-M)_{0}=Ks_{\rm RC,intr.}-M_{Ks}+\Delta M_{K},
\end{equation}
where $Ks_{\rm RC,intr.}$ is the intrinsic, reddening free magnitude
of RC stars, $M_{Ks}$ is their absolute $Ks$-magnitude, and $\Delta
M_{K}$ the population correction for $M_{Ks}$. We adopt the same
values as \cite{Nishiyama:2006ai} for $\Delta M_{K}=-0.07$ and use
$M_{Ks}=-1.54$ (see above). By assuming a GC distance, we can use the
above equation to estimate the reddening free magnitude of RC
stars. By comparing it subsequently with the magnitude of the RC clump
peak as inferred from the luminosity function, we can then obtain an
estimate of the absolute extinction in $Ks$ toward the GC.

\begin{figure*}[!htb]
\includegraphics[width=\textwidth]{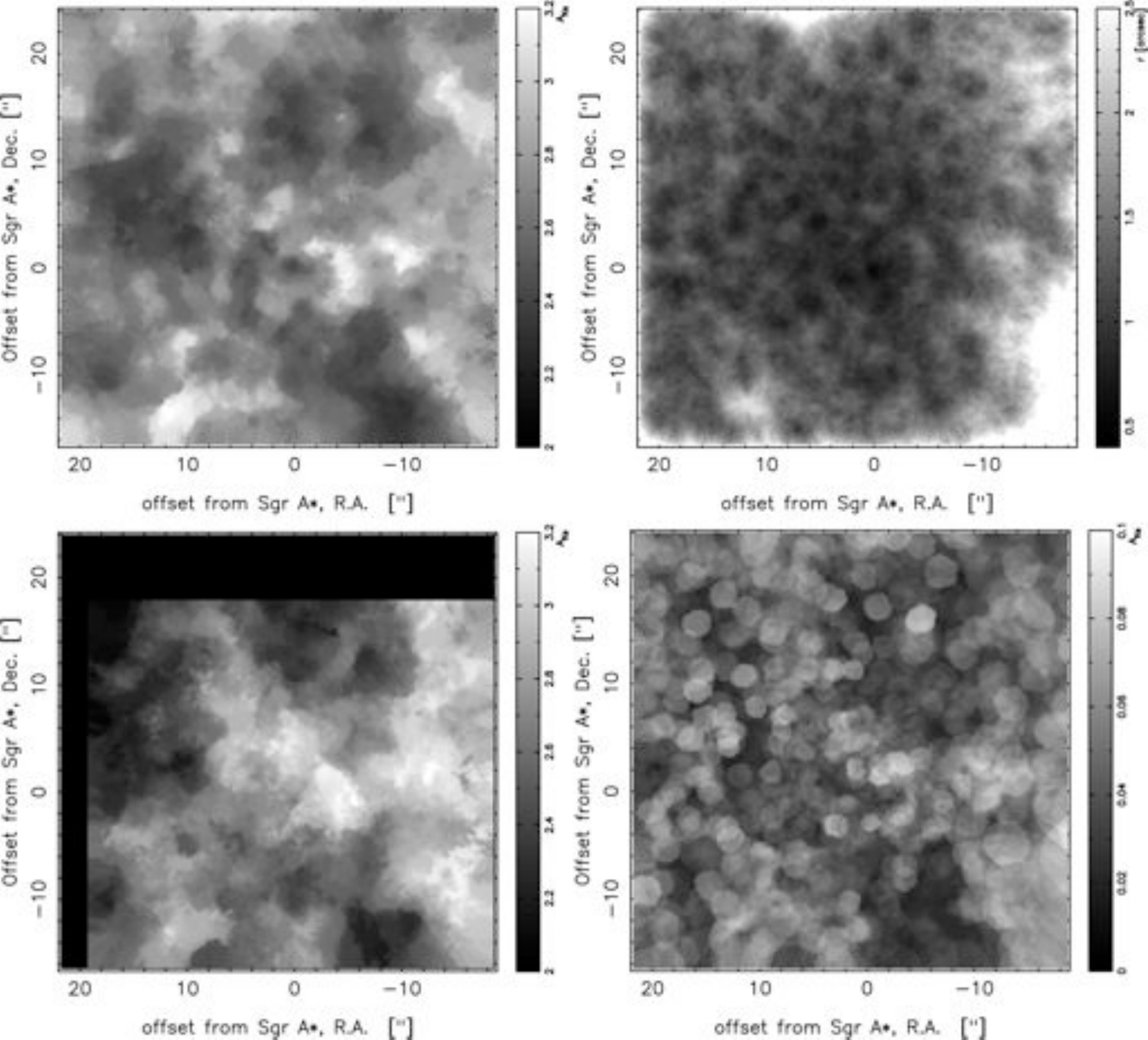}
\caption{\label{Fig:extmap} Top left: Map of $A_{Ks}$ for the GC
  derived from $H-Ks$ colors of all stars detected in these bands,
  after exclusion of fore- and background stars. Top
  right: Map of the resolution of the extinction, i.e. the radius in
  arcseconds at each position within which the 20 stars for the given
  measurement were detected. Bottom left:  Map of $A_{Ks}$ for the GC
  derived from $Ks-L'$ colors.  Bottom right: Map of the statistical
  uncertainty of $A_{Ks}$, corresponding to the extinction map shown
  top left.}
\end{figure*}

Numerous efforts to determine the distance of the GC have been made
recently (the cited uncertainties are the quadratically combined
statistical and systematic uncertainties given in the respective
works): $7.52\pm0.36$\,kpc \citep[bulge red clump
stars,][]{Nishiyama:2006ai}, $8.33\pm0.35$\,kpc \citep[stellar orbit
around Sagittarius\,A*,][]{Gillessen:2009qe}, $8.0\pm0.6$\,kpc
\citep[stellar orbit around Sagittarius\,A*,][]{Ghez:2008oq},
$7.94\pm0.45$\,kpc \citep[Population II Cepheids and RR Lyrae
stars][]{Groenewegen:2008by}, $8.07\pm0.35$\,kpc \citep[proper motion
and line-of-sight velocity dispersion of late-type stars in central
parsec,][]{Trippe:2008it}, $8.4\pm0.6$\,kpc \citep[trigonometric
parallaxes of masers measured with VLBI,][]{Reid:2009nx},
$8.24\pm0.43$\,kpc \citep[Mira stars,][]{Matsunaga:2009qp}, and
$7.9\pm0.8$\,kpc \citep[trigonometric parallax of Sgr B2 measured with
VLBI,][]{Reid:2009eu} . These distance estimates are based on
different data sets and different methods and can probably be
considered largely independent measurements. It appears therefore
justified to combine these measurements to a weighted mean of
$R_{0}=8.03\pm0.15$\,kpc. This corresponds to a distance modulus of
$14.52\pm0.04$.  This simple exercise demonstrates that -- under the
assumption that these measurements are indeed statistically
independent -- $R_{0}$ is already known with the surprisingly small
uncertainty of just $2\%$.  Using this value we obtain a reddening
free magnitude of the RC bump at the GC of $Ks_{\rm RC,
  intr.}=13.05\pm0.09$. All uncertainties have been quadratically
combined ($\Delta (m-M)_{0}=0.04\,$mag, $\Delta M_{Ks}=0.04\,$mag,
$\Delta\Delta M_{K}=-0.07\,$mag).

The fit of the RC bump in the $Ks$LF with a Gaussian (plus an
exponential to fit the underlying luminosity function) results in
$Ks_{\rm RC}=15.59\pm0.07$. Here, the uncertainties from the fit of
the RC peak and from the absolute photometric calibration have been
combined quadratically. From the difference with the theoretical
reddening free magnitude of the RC bump, this results in an average
extinction of $A_{Ks}=2.54\pm0.12$ toward the GC.

Assuming $H-Ks=0.07\pm0.03$ and $Ks-L'=0.07\pm0.03$ for the RC stars,
we obtain in the same way estimates for $A_{H}=4.48\pm0.13$ and
$A_{L'}=1.27\pm0.18$.  Note that these values depend only on the
well-known properties of RC stars and on the GC distance. They are
independent of any assumptions on the extinction-law.

Assuming the validity of a power-law for the extinction law between
$H$ and $Ks$ and between $Ks$ and $L'$, respectively \citep[see,
e.g.,][]{Nishiyama:2009oj}, we can derive from the above measurements
the respective power-law indices $\alpha_{H-Ks}=2.21\pm0.24$ and
$\alpha_{Ks-L'}=1.34\pm0.29$. Here the uncertainties include the
uncertainties of the effective wavelengths at the corresponding
bands. The effective wavelengths adopted in this work are
$\lambda_{\rm eff, H}=1.677\pm0.018\,\mu$m, $\lambda_{\rm eff,
  Ks}=2.168\,\pm0.012\,\mu$m, and $\lambda_{\rm eff,
  L'}=3.636\,\pm0.012\,\mu$m. Details on how these effective
wavelengths have been calculated are given in
appendix\,\ref{app:efflambda}.

As can be seen in the color-magnitude diagram shown in
Fig.\,\ref{Fig:colmag} (top left) there is a clear dependency of the
$Ks$ peak magnitude of the RC feature on $H-Ks$. Because of the narrow
intrinsic distribution of colors and magnitudes of the RC stars we
expect this trend to be almost exclusively due to differential
extinction. In order to quantify this relation, we extracted the
$Ks$LF in narrow ranges of $H-Ks$. The peaks of the RC features were
then fitted with a Gaussian. This method is very similar to the one
applied by \citet{Nishiyama:2006tx}, who examined the location of the
RC feature in different fields toward the GC in order to derive the
reddening law.  Figure\,\ref{Fig:RCfit} shows the thus measured RC
$Ks$ peak magnitude vs.\ $H-Ks$ along with a linear fit. Assuming
$A_{\lambda}\propto \lambda^{-\alpha_{H-Ks}}$, the slope of the line
gives $\alpha_{H-Ks}=2.54\pm0.15$.  This value agrees within the
uncertainties with the one derived from the RC bump in the
$H$ and $Ks$ luminosity functions (see above) and thus
serves as a cross-check. However, we consider the value
$\alpha_{H-Ks}=2.21\pm0.24$ derived from the RC as more reliable
(see discussion section).

The $Ks$ vs. $H-Ks$ color magnitude diagram shows a decreasing
completeness with increasing reddening. This is reflected in
Fig.\,\ref{Fig:colmag} by the increasing faint magnitude limit of the
cloud of dots toward redder colors. We have seen above that
incompleteness does not significantly affect the measured peak of the
RC in the $Ks$LF.  We have not used any completeness correction for
the fit of the RC peak in the color magnitude diagram, but tested its
reliability by omitting the two ``reddest'', i.e. highest $H-Ks$, data
points. The fit then results in $\alpha_{H-Ks}=2.71\pm0.18$. If we
additionally omit the bluest data point from the fit, which may be an
outlier, we obtain $\alpha_{H-Ks}=2.45\pm0.14$. We conclude that
incompleteness due to reddening does not have any significant
influence on the value of $\alpha_{H-Ks}$ derived from the
color-magnitude diagram in Fig.\,.\ref{Fig:RCfit}.

As can be seen in the upper right panel of Fig.\,\ref{Fig:colmag},
there is no clear trend of the RC visible in the plot of $L'$
vs.\ $Ks-L'$. This is mainly because the number of stars measured in
$L'$ is smaller (smaller FOV) and because the range spanned by the
$Ks-L'$ colors is limited due to the weak extinction in $L'$ . We can
therefore not apply this method to determine $\alpha_{Ks-L'}$ (see
below for a different method, however).

\section{An extinction map for the GC \label{sec:extmap}}

We produced an extinction map for the FOV of the $H$ and $Ks$
observations of the GC by using the median of the $H-Ks$ colors of
the 20 nearest stars at each position.  For this purpose we used
  all stars from our list, but excluded extremely blue ($H-Ks<1.8$) or
  red ($H-Ks>2.8$) ones as fore- or background stars. The intrinsic colors
of almost all stellar types observable at the GC are similar at these
wavelengths. Nevertheless, we tried to improve the estimates
by applying a 0th order correction for different intrinsic stellar colors.
Stars fainter than $Ks=14.5$ were assigned $(H-Ks)_{0}=0.07$ and stars
brighter than this value were assigned $(H-Ks)_{0}=0.2$. Known
early-type stars \citep[taken from][]{Buchholz:2009sp} were assigned
$(H-Ks)_{0}=-0.03$. In order to assign these intrinsic colors, we used
the GC distance, approximate extinction, and the magnitude of the stars to roughly
guess their type \citep[see, e.g., Table\,3 in][]{Buchholz:2009sp} and
assigned colors from tables\,7.6 and 7.7 in \citet{Cox:2000yb}.

In order to check the importance of distinguishing different intrinsic
stellar colors we also created a map of $H-Ks$ assuming just one
uniform intrinsic color, $H-Ks=0.07$ for all stars.  The resulting map
looks almost identical to the one including the more refined
corrections (the mean difference between the two maps is
$\Delta(H-Ks)=0.02\pm0.02$). We conclude that it would be, in
principle, acceptable to assume just one common intrinsic color for
our entire sample of stars because only $\sim16\%$ of our sample of
stars are brighter than $Ks=14.5$. The majority are, in fact, red
clump stars. Although this shows that distinguishing different types
of stars in the 0th order color correction is almost negligible, we
nevertheless include this correction in order to avoid bias in regions
where certain types of stars may be crowded. For example, the majority
of stars within $0.5"$ of Sgr\,A* are of early-type
\citep[e.g.][]{Eisenhauer:2005vl}.

The map of color excesses, $E(H-Ks)$, was converted to an
extinction map by assuming a power-law extinction-law with an index
$\alpha_{H-Ks}=2.21\pm0.24$ for the effective wavelengths of
$\lambda_{H}=1.677\pm0.018\,\mu$m and
$\lambda_{Ks}=2.168\pm0.012\,\mu$m. The uncertainty of these three
parameters combined with the uncertainty of the photometric zero
points and of the intrinsic stellar colors results in a combined
systematic uncertainty of $\sim9\%$ on the calculated values of
$A_{Ks}$. The extinction map based on $H-Ks$ colors is shown in the
top left panel of Fig.\,\ref{Fig:extmap}. The top right panel of this
figure shows the resolution of the extinction map, i.e.\ the radius at
each position within which 20 stars were detected. The
  statistical uncertainty, $RMS/\sqrt{N-1}$ (with $N=20$), of the
  extinction at each point in the map is shown in the bottom right
panel of Fig.\,\ref{Fig:extmap}. It is $<0.1$\,mag for the entire map.
As a cross-check, we also created an extinction map based on $Ks-L'$
colors, in a completely analogous way. It is presented in the bottom
left panel of Fig.\,\ref{Fig:extmap} and shows the same patterns as
the extinction map based on $H-Ks$ colors. Note that details will be
more reliable in the map based on $H-Ks$ because of the higher point
source density measured in these bands. Based on the assumption that
the extinction law between $Ks$ and $L'$ is a power-law, we determined
the corresponding power-law index, $\alpha_{Ks-L'}$, by requiring that
the average $A_{Ks}$ must be the same, based on $H-Ks$ or $Ks-L'$
colors. A value of $\alpha_{Ks-L'}=1.35$ has been obtained, with a
statistical uncertainty of $0.12$ and a systematic uncertainty -- due
to uncertainties in the effective wavelengths and overall photometric
calibration -- of $0.29$. This value of $\alpha_{Ks-L'}$ agrees
  very well with the one derived from the location of the RC bump in
  the $Ks$ and $L'$ luminosity functions.

\begin{figure}[!htb]
\includegraphics[width=\columnwidth]{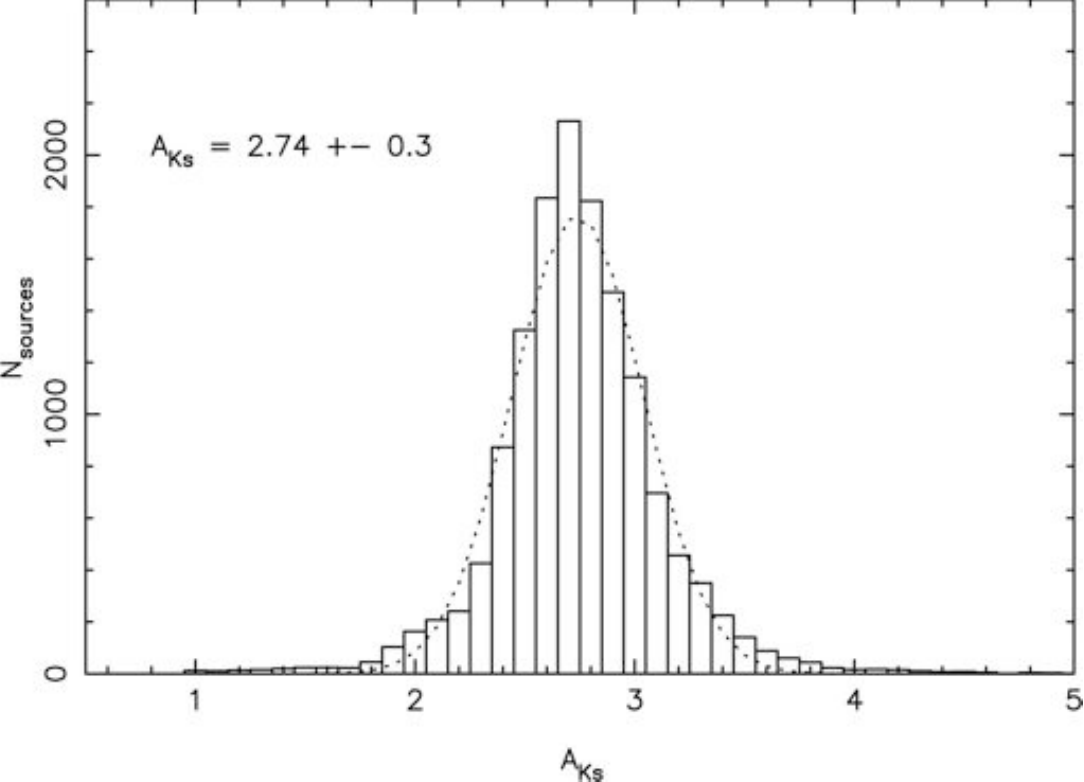}
\caption{\label{Fig:exthist} Histogram of the extinction,
  $A_{Ks}$, based on $H-Ks$ colors.}
\end{figure}

A histogram of the values of $A_{Ks}$ measured based on the $H-Ks$
colors is shown in Fig.\,\ref{Fig:exthist}. The histogram is fit well
by a Gaussian with a mean value of $A_{Ks,mean}=2.74$\,mag and a
standard deviation of $0.30$\,mag.

\begin{figure}[!htb]
\includegraphics[width=\columnwidth]{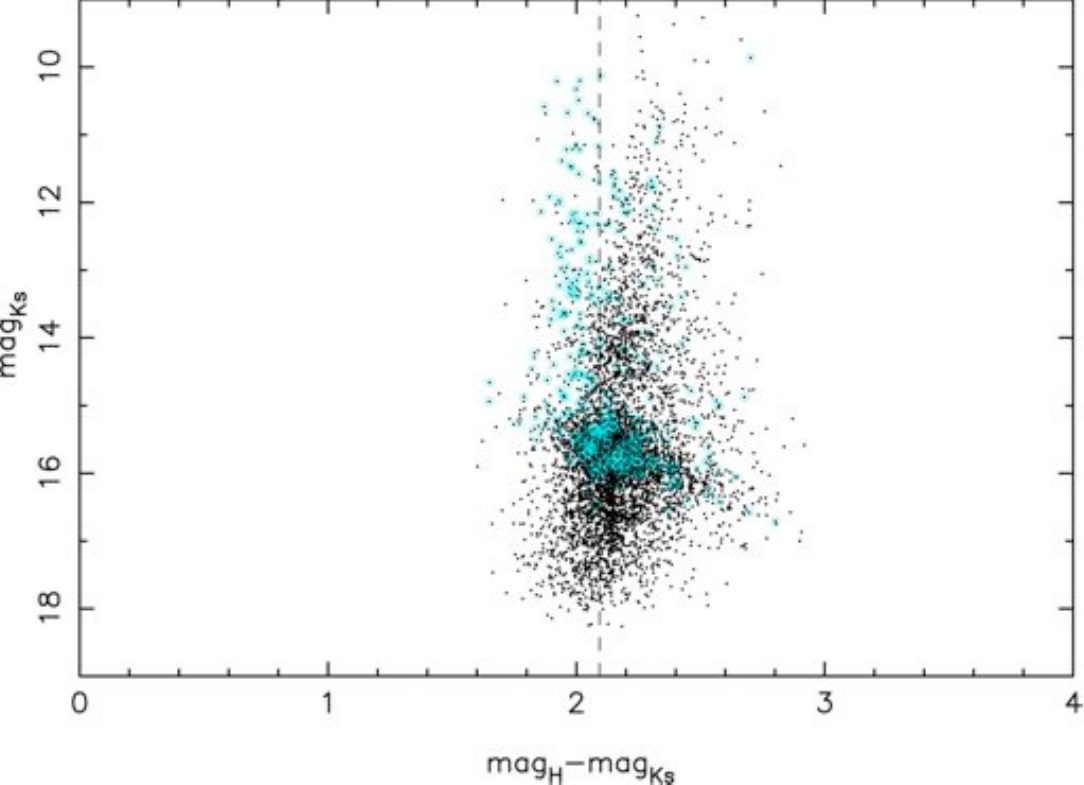}
\caption{\label{Fig:hkcorr} Differential extinction corrected
  color-magnitude diagram for the GC (assumed mean extinction is
  $A_{Ks}=2.74$). The dashed line indicates the mean $H-Ks$
  color. Known early-type stars are marked by blue circles.}
\end{figure}

The extinction map was used to produce a differential extinction
corrected color magnitude diagram of $Ks$ vs.\ $H-Ks$. It is shown in
Fig.\,\ref{Fig:hkcorr}, after exluding all stars with $H-Ks<1.8$ as
foreground and $H-Ks>2.8$ as background stars.  It can be seen how the
extinction correction significantly reduces the scatter in the diagram
(compare with the uncorrected diagram in the upper left panel of
Fig.\,\ref{Fig:colmag}). The red clump is very well defined and the
giant sequence can be perceived clearly to the right of the dashed
line, which indicates the median $H-Ks$ color. Stars on the left of
the dashed line are expected to be either foreground or early-type
stars. Known early-type stars \citep{Buchholz:2009sp} are marked with
blue circles. They provide a sanity check on the reliability of
identifying early-type stars via extinction corrected $H-Ks$ colors.
At $Ks<14$ about $65\%$ of the known early-type stars are identified
correctly. How high is the probability to misidentify late-type stars
as early-type with this method? Since the FOV of our work is larger
than in comparable work, we limit the analysis for investigating this
question to the region within $12"$ of Sgr\,A* and to magnitudes
brighter than $Ks=14$. With these limits we can reasonably assume that
the number of known and correctly classified stars (as early- or
late-type) from literature is close to $100\%$.  Within $12"$ of
Sgr\,A* we find that about $30\%$ of the stars brighter than $Ks=14$
identified by the extinction corrected $H-Ks$ colors may be erroneous
identifications. Potential important sources of error are, for
example, photometric errors and inaccuracies of the extinction map
because of variability at scales below the resolution of our
extinction map or because of locally different extinction related
possibly with individual sources, e.g.\ due to circumstellar material.
Nevertheless, considering the crudenes of the method presented here, a
success rate of $\sim2/3$ appears reasonable and suggests that our
method may be useful for large-scale surveys in order to identify
candidates for massive, young stars in highly extincted areas. At
magnitudes $Ks>14$ the extinction corrected $H-Ks$ criterion is too
crude and does not allow to separate early- and late-type stars in our
sample. However, future work may provide more accurate colors and
extinction values.

We have marked all stars with $Ks<14$ and a color bluer than the
median $H-Ks$ (after excluding foreground stars, see above) with
crosses in the $Ks$ image shown in Fig,\,\ref{Fig:mosaicKs}. Many
early-type stars reported by
\citet{Paumard:2006xd,Buchholz:2009sp,Do:2009tg} and other authors are
correctly identified. Following the error analysis in the
  preceding paragraph of the order $30\%$ of the stars may be
spurious identifications and a similar number of early-type stars may
be missed by this method. Nevertheless, this is the first time that
early-type stars at the GC can be identified via broad-band
imaging. Even taking into account the large estimated uncertainties,
this rather crude method confirms the finding of
\citet{Buchholz:2009sp} that there exists a population of early-type
stars beyond a projected radius of $0.5$\,pc from Sagittarius\,A*.

\begin{figure}[!htb]
\includegraphics[width=\columnwidth]{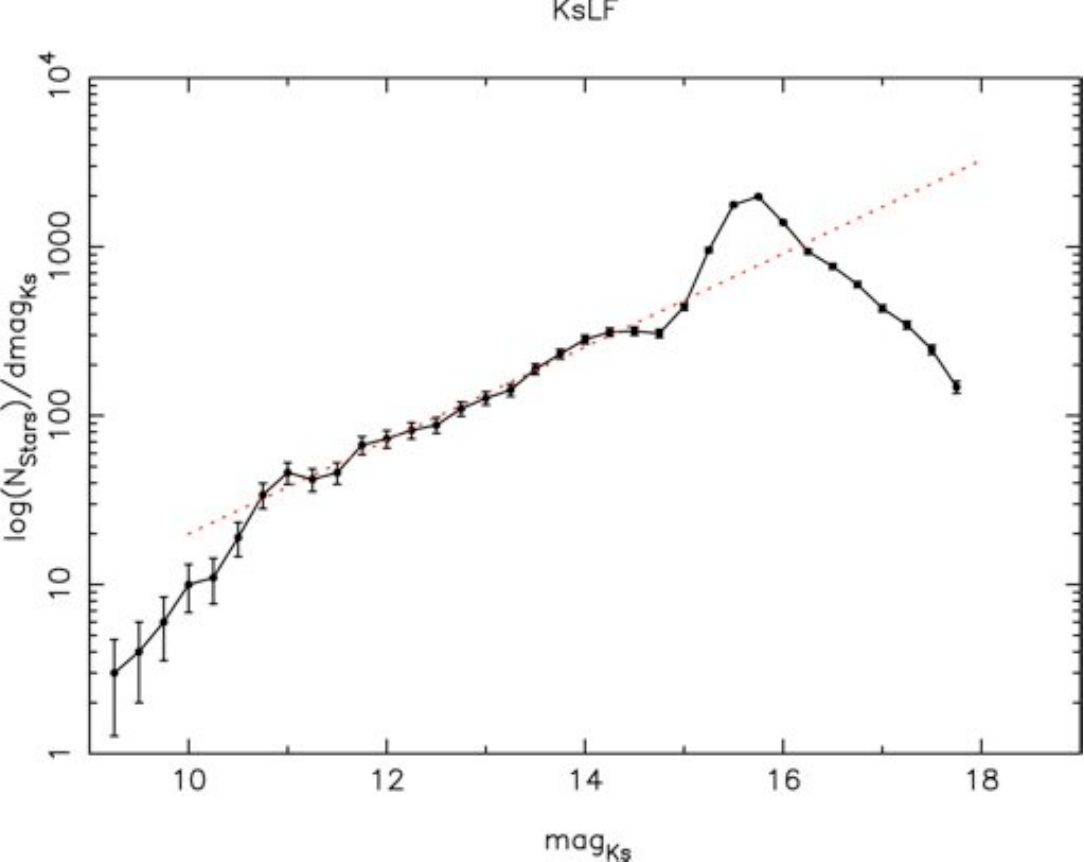}
\caption{\label{Fig:LFcorr}  $Ks$ luminosity
  function for the GC, fore- and background stars excluded, and
  corrected for differential extinction. The mean extinction has been
  set to $A_{Ks}=2.74$.  This
  luminosity function has not been corrected for completenes. The red
  dashed line is a power law fit to the stars $11<Ks<14$. It has a
  power-law index of $0.27\pm0.02$.}
\end{figure}

Finally, we have used the extinction map to produce a$Ks$
  luminosity function corrected for differential extinction (fore- and
  background stars excluded, mean extinction $A_{Ks}=2.74$, no
  incompleteness correction applied). It is shown in
Fig.\,\ref{Fig:LFcorr}. By comparing the power-law fit (dashed line)
with the data, one can estimate that in the shallow observations used
in this work (in order to avoid saturating the brightest stars) the
completeness drops below 50\% already at $Ks\approx17$ (see also
Table\,\ref{Tab:completeness}).

\section{Discussion}

For comparison of the derived photometry and extinction with other
work, it is important to refer to an effective wavelength, i.e.\ to
take into account such factors as the SED of the stars, extinction,
and the filter transmission curves. The measured value of the exponent of the
power-law extinction-law is strongly dependent on the effective
wavelength \citep[see, e.g.,][]{Nishiyama:2006tx,Gosling:2009kl,Stead:2009oq}.  The
transmission curves of the NACO filters are available at the ESO
website. In order to facilitate comparison of our work, we have
calculated the effective wavelengths for the photometry presented
here, using equation (A3) of \citet{Tokunaga:2005jw}.

We have found a significantly steeper extinction law at NIR
wavelengths than in earlier work \citep[e.g.,][see the latter paper
for a review ]{Rieke:1985fq,Draine:1989eq,Mathis:1990uf}. There
exists, however, very good agreement between the values for
$\alpha_{H-Ks}=2.21\pm0.24$ (and $\alpha_{H-Ks}=2.54\pm0.15$,
respectively) and $\alpha_{Ks-L'}=1.34\pm0.29$ (and
$\alpha_{Ks-L'}=1.35\pm0.31$, respectively) found in this work and the
respective values from other recent research. \citet{Gosling:2009kl}
report $\alpha_{J-Ks}=2.64\pm0.52$ based on NIR observations of the
nuclear bulge. Note that they find variability of this power-law index
along the line-of-sight, on scales as small as $5"$. We have not
examined spatial variability of the power-law index in our work. Since
we do not have $J$-band observations available and since the
  extinction law appears to be clearly different between $J$ and $Ks$
  and $Ks$ and $L'$, we cannot determine the extinction-law toward
individual stars or small groups of stars like \citet{Gosling:2009kl}.
\citet{Nishiyama:2009oj} find $\alpha_{H-Ks}=2.0$ and
$\alpha_{Ks-3.6}=1.37$, based on NIR observations of the Galactic
nuclear bulge. \citet{Stead:2009oq} examined 8 regions from the UKIDDS
Galactic Plane Survey and report $\alpha_{\rm NIR}=2.14\pm0.05$.
These values agree with our results within the $1\sigma$ - combined
statistical and systematic - uncertainties. An exception is the value
$\alpha_{H-Ks}=2.54\pm0.15$ determined from the color dependence of
the RC in the $H-Ks$ vs.\ $Ks$ color-magnitude diagram. Nevertheless,
this error lies within the $1\sigma$ uncertainty of the value
$\alpha_{H-Ks}=2.21\pm0.24$ that we have derived in this work from the
RC positions in the $Ks$ and $H$ luminosity functions. We consider the
latter value more trustworthy because it is more straightforward to
determine (no cutting out of regions from the color-magnitude
diagram), is based on large numbers of stars, and agrees very well with
the recent works cited above.

The extinction map based on $H-Ks$ colors from this work
resembles closely the one presented in \citet{Buchholz:2009sp}. The
overall extinction we find in the present work is, however, lower than the
one reported by \citet{Buchholz:2009sp} -- $A_{Ks}=2.74\pm0.30$
compared to $A_{Ks}=3.14\pm0.43$. The reason for this difference is
the assumption of a flatter extinction law \citep[based
on][]{Draine:1989eq} in \citet{Buchholz:2009sp}. The extinction map
presented by \citet{Schodel:2007tw} is also similar to the one
presented here, but shows a smaller FOV and is also based on a
different extinction law. Agreement can also be seen between the
extinction map produced in this work and the one presented by
\citet{Scoville:2003la}. The map in the latter paper encompasses a
much larger region, but is less homogeneous than the one produced here
because it is derived from the line emission of ionized gas. Due to
the strong radio emission from Sagittarius\,A* the latter authors
could not reliably constrain the extinction in the immediate
surroundings of Sagittarius\,A*.  \citet{Scoville:2003la} used an
extinction-law of the form $A_{\lambda}\propto\lambda^{-1.6}$,
significantly different from what is found here and in other recent
work on this subject.

A caveat may apply to the regions with the strongest extinction.
As can be seen in Fig.\,\ref{Fig:mosaicKs}, there are some regions
(e.g.\ at an offset of $\sim+12"$\,E and $\sim-13"$\,S of Sagittarius\,A*)
with very high extinction and consequently few stars. Extinction may
be under-estimated by our method in these patches of extremely high
extinction because the number of stars that can be detected behind a
strong extinction screen may  be very small. 

Finally, useful for analyses of the so-called ``S-cluster'' of stars
that are tightly bound to Sagittarius\,A* and of the NIR emission of
Sgr\,A* itself, we can use the extinction map presented in
Fig.\,\ref{Fig:extmap} to calculate the mean extinction and its
standard deviation toward Sagittarius\,A*. For a circular region with
a radius of $0.5"$ centered on Sagittarius\,A* we obtain $A_{H,
  SgrA*}=4.35\pm0.12$, $A_{Ks, SgrA*}=2.46\pm0.03$, and $A_{L',
  SgrA*}=1.23\pm0.08$. The uncertainty of $A_{Ks, SgrA*}$
  corresponds to the standard deviation of extinction within the
  circular aperture.  The uncertainties for $A_{H, SgrA*}$ and
$A_{L', SgrA*}$ include additionally the uncertainties of $\alpha_{H-Ks}$ and
$\alpha_{Ks-L'}$. These extinction values are different from the
  ones that have typically been used in the analysis of NIR
lightcurves from Sgr\,A*, e.g., $A_{H}=4.3$, $A_{K}=2.8$, $A_{L'}=1.8$
\citep{Genzel:2003hc,Eckart:2006sp}, $A_{K}=3.3$ \citep{Do:2009ij}, or
$A_{L'}=1.8$ \citep{Ghez:2004fx}. While the impact on
emission/accretion models of Sgr\,A* will probably be minor, we point
out that the new extinction values at $Ks$ and $L'$ imply that the
de-reddened emission from Sgr\,A* may be weaker than assumed by a
factor $1.4-2.2$ at $Ks$ and a factor of $\sim1.7$ at $L'$.

A correct estimate of the extinction is crucial to address one of the
long standing issues in the central parsec, namely the apparent
incompatibility between the ionizing flux arising from the massive
stars and the ionisation of the gas
\citep{Najarro:1997qe,Lutz:1999yf}. Further, these authors found that
the population of HeI stars, displaying fairly large luminosities was
unusually high.  Such incompatibility was recently claimed to be
solved by \citet{Martins:2007sf} with the inclusion of line-blanketing
in the models, leading to lower bolometric luminosities and lower UV
fluxes for the HeI stars.  However, we attribute most of the
discrepancy between the results of \citet{Najarro:1997qe} and
\citet{Martins:2007sf} for the HeI stars to the estimate of $A_{K}$
between both investigations. Indeed, the final $A_{K}$ values computed
by \citet{Najarro:1997qe} were 50\% larger than the ones adopted by
\citet{Martins:2007sf} from the extinction map of
\citet{Schodel:2007tw}. Such difference was translated into roughly
one magnitude in luminosity and was responsible for the much lower
luminosity of the HeI stars revised by \citet{Martins:2007sf}.
Similarly, the nature of the "S" stars is also affected. While
\citet{Martins:2008fe} adopted $A_{K}=2.25$ for S2 it turns out that
the extinction map they used was calibrated on S2 with an $A_{K} =
2.8$ \citet{Schodel:2007tw}.  That would have changed S2 from being a
dwarf to being a giant. It turns out that the actual value of
$A_{Ks}=2.46$ toward S2 is not too different from the value assumed
(erroneously) by \citet{Martins:2008fe}.  As we show in this work, and
as other similar recent investigation has shown, the solution and key
to these problems is that the power-law slope of the NIR extinction
law is significantly steeper than what was assumed in earlier work. This
leads to lower extinction in the $K$-band, but significantly stronger
extinction at shorter wavelengths.

\begin{figure}[!htb]
\includegraphics[width=\columnwidth]{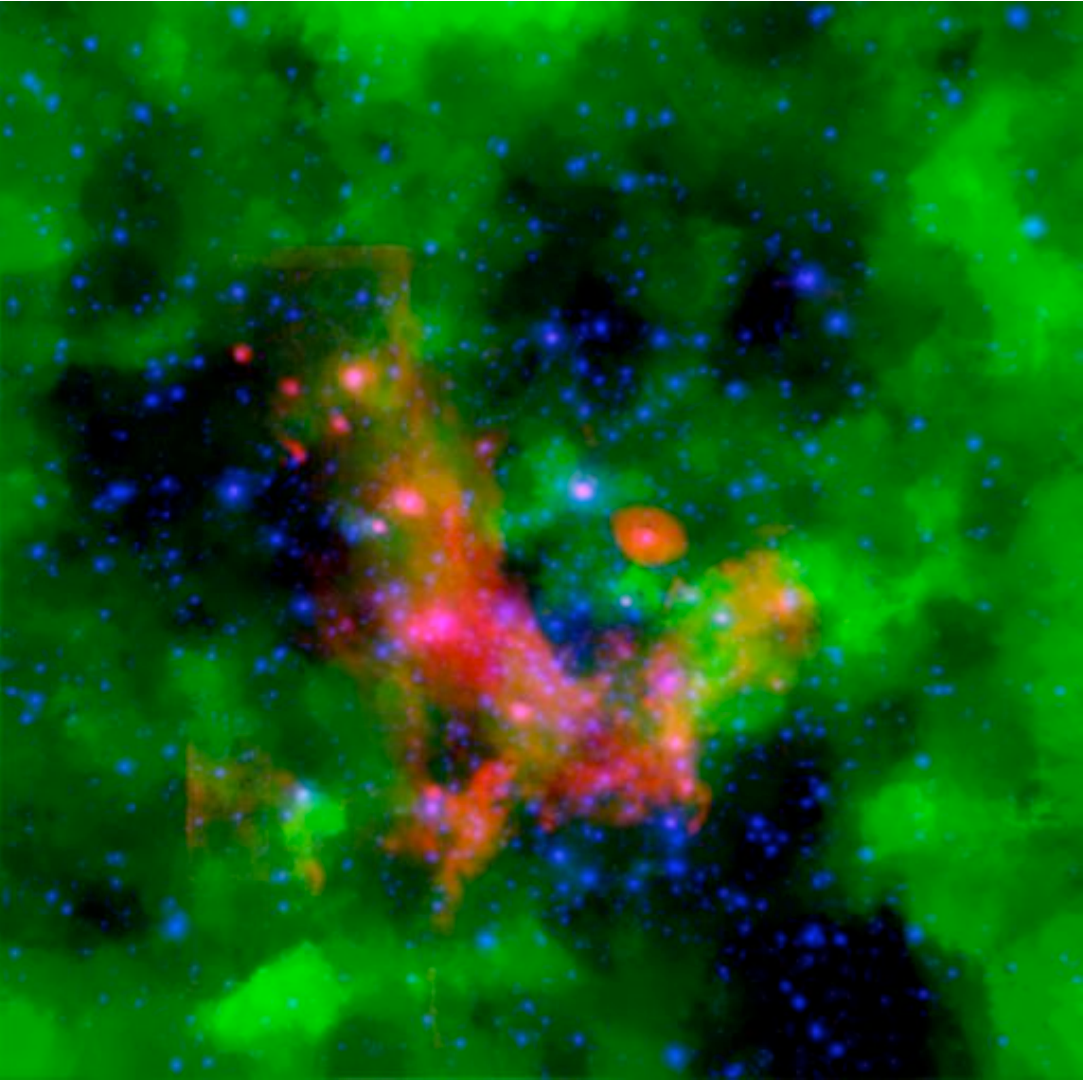}
\caption{\label{Fig:composite}  Ilustration of the relation between
  the stars detected at $Ks$ (blue), the $A_{Ks}$ extinction (green),
  and the diffuse emission in the mid-infrared (PAH\,I, red).}
\end{figure}

In Fig.\,\ref{Fig:composite} we show a color composite image of the
stars, detected at $Ks$, the $A_{Ks}$ extinction, and the MIR (PAH\,I)
emission \citep[data from][]{Schodel:2007hh}. It can be seen how the
extinction is low in a region running roughly from SW to NE. This
creates an apparent high stellar density and therefore elongation of
the NSC along the galactic plane \citep[as was also shown
by][]{Schodel:2007tw}. Similarly, the lower extinction to the NW of
Sgr\,A* is responsible for the higher stellar density observed in that
region. The strong influence that extinction has on the appearance of
the nuclear star cluster can also be nicely seen in  Fig.\,1 of \citet{Scoville:2003la}. 
 There appears to be no obvious relation between the warm dust
emitting in the mid-infrared and the extinction. This is no surprise
because the surface density of HI and $H_{2}$ and the warm dust
associated with it in the mini-spiral is just of the order a few
$\times10^{4}$\,cm$^{-3}$\citep[see, e.g., discussion of this topic
in][]{Moultaka:2009ul}. The extinction toward the GC is largely
produced by clouds in the foreground, with the possible exception of
some dust-enshrouded individual objects \citep{Moultaka:2009ul} .

For our analysis we have used a weighted average of recent
measurements of the distance to the GC, assuming that all those
measurements are independent. We did not investigate this assumption
in depth. However, we took into account the systematic uncertainties
of the individual measurements when computing the weighted average. As
a further simple test we can use sub-samples of the measurements and
check the resulting values of $R_{0}$ for consistency. Using only
geometrical estimates
\citep{Ghez:2008oq,Trippe:2008it,Gillessen:2009qe,Reid:2009nx,Reid:2009eu}
we obtain $R_{0,geom,1} = 8.20\pm0.22$\,kpc \citep[emitting the
estimate from the S0-2/S2 orbit by][]{Ghez:2008oq} and $R_{0,geom,2} =
8.10\pm0.26$\,kpc \citep[emitting the estimate from the S0-2/S2 orbit
by][]{Gillessen:2009qe}. Using only measurements based on variable and
RC stars, we obtain $R_{0,stars} = 7.85\pm0.24$\,kpc.  As an
additional test we computed the unweighted average of all
measurements: $R_{0,unweighted}=8.05\pm0.28$\,kpc. It agrees within
the uncertainties with the weighted average. If we remove the
measurement with the strongest apparent deviation
\citep[$7.52\pm0.36$\,kpc,][]{Nishiyama:2006ai}, we obtain
$R_{0,unweighted}=8.12\pm0.20$\,kpc. The unweighted averages of the
sub-sets of distance measurements considered above agree also well
with the weighted results. We therefore believe that the value for
$R_{0}= 8.03\pm0.15$\,kpc derived in this work and its related
uncertainty can be considered reliable.

We believe that future work should focus on deriving the power-law
index for the NIR extinction-law toward the GC with higher precision
and examine its variability over the central parsecs. This could be
achieved, for example, by deep AO imaging using a large number of
narrow band filters across the near-infrared. The main limitation for
this kind of work may come from the still limited performance of AO
systems at short NIR wavelengths ($J$-band), combined with the long
exposure times required by the high extinction. An additional great
difficulty, in fact the main source of uncertainty in this work, is
the difficulty to obtain precise and accurately calibrated photometry
from AO data (crowding, large seeing foot of PSF, spatial variability
of PSF).  We have, however, no doubt that there will be significant
technical progress in AO techniques and related data processing
software in the next few years.

\section{Summary}

We provide photometry for $\sim7700$ point sources extracted from
adaptive optics observations of the central parsec of the Galactic
center in the $Ks$, and for many of them also in the $H$ and $L'$
  bands. The data are made publicly available as electronic
  on-line material. Adaptive optics photometry has been published
previously in some papers. However, we believe that the data presented
in this work supersede those earlier results or are complementary to
them because of reasons such as: (a) limited field-of-view of previous
observations or focus on the central few 0.1\,pc; (b) isoplanatic
effects were not taken properly into account; (c) bias by the
application of methods such as Lucy-Richardson deconvolution (the
companion paper to this work, by Schoedel, A\&A, 2009, accepted for
publication, explains our photometric method and points out the bias
introduced by the Lucy-Richardson method); (d) no transparent
explanation of the photometric calibration. This data set is the most
comprehensive one published so far.

Based on the photometric results, we analyze extinction toward the
central parsec of the GC. Absolute values of extinction in the three
examined bands are derived based on the well-known properties of red
clump stars, without the need to measure stellar colors and without
having to assume an extinction law. The values obtained are
$A_{H}=4.48\pm0.13$\,mag, $A_{Ks}=2.54\pm0.12$\,mag, and
$A_{L'}=1.27\pm0.18$\,mag. This results in
$A_{H}:A_{Ks}:A_{L'}=1.76:1:0.50$. Assuming the validity of a
power-law extinction law between $H$ and $Ks$ and between $Ks$ and
$L'$, this is equivalent to power-law indices of
$\alpha_{H-Ks}=2.21\pm0.24$ and $\alpha_{Ks-L'}=1.34\pm0.29$. This is
significantly steeper than in earlier work
\citep[e.g.][]{Rieke:1985fq,Draine:1989eq}, but agrees with recent
studies of extinction toward the nuclear bulge
\citep{Nishiyama:2009oj,Gosling:2009kl} and other targets in the
Galactic plane \citep{Stead:2009oq}. We also confirm the flattening of
the extinction law beyond $\sim3\,\mu$m \citep[see
also][]{Lutz:1996oz,Indebetouw:2005rp,Viehmann:2005uk,Flaherty:2007if,Nishiyama:2009oj}.

Based on $H-Ks$ colors of the stars, we provide a detailed
extinction map for a FOV of $40"\times40"$ ($\sim1.5\,{\rm
  pc}\times1.5\,{\rm pc}$) centered roughly on Sagittarius\,A*. The
spatial resolution of the map is $\sim1"$. Both statistical and
systematic uncertainties of the extinction map are smaller than
$10\%$. The extinction map is made publicaly available as
  electronic on-line material.  We find that extinction varies
significantly on arcsecond scales. The mean extinction based on
stellar colors is found to be $A_{Ks}=2.74\pm0.30$\,mag, in good
agreement with the value of mean extinction based on the RC method. An
extinction map based on $Ks-L'$ colors, although less accurate because
of lower point source surface density and possible contamination by
diffuse emission from dust in the mini-spiral, confirms the patterns
seen in the extinction map based on $H-Ks$ colors. Requiring the mean
extinction between the two maps to be the same, we derive
$\alpha_{Ks-L'}=1.35\pm0.31$, in excellent agreement with the value of
this parameter derived from the mean magnitudes of the RC bump (see
above).

Applying the derived extinction map to the observed $H-Ks$ colors
  of the stars, we are able to correct the corresponding
color-magnitude diagram. This reduces the scatter in the diagram
considerably and makes it possible to distinguish between stars in the giant
branch and young,  massive stars. We confirm recent findings of the presence of
of young, massive stars at projected distances $>0.5$\,pc from
Sagittarius\,A*.

A minor, but very interesting by-product of this work is a weighted
average of recent (years 2006-2009) estimates of $R_{0}$. We find that
the distance of the GC is constrained with an uncertainty  of just $2\%$:
$R_{0}=8.03\pm0.15\,$kpc.

 For a circular region with a radius of $0.5"$ centered on
 Sagittarius\,A* we obtain $A_{H, SgrA*}=4.35\pm0.12$, $A_{Ks,
  SgrA*}=2.46\pm0.03$, and $A_{L', SgrA*}=1.23\pm0.08$. These values
imply that the NIR emission from Sgr\,A* in the $Ks$ and $L'$-bands
may be weaker than previously assumed, by factors $1.4-2.2$. 

We have used the following effective wavelengths in this work:
$\lambda_{H}=1.677\pm0.018\,\mu$m, $\lambda_{Ks}=2.168\pm0.012\,\mu$m,
and $\lambda_{\rm eff, L'}=3.636\,\pm0.012\,\mu$m. Note that the
effective wavelength is a complex function of the transmission
functions of the atmosphere, the filter, the optical system,
extinction, and stellar type. For discussions of this problem,
see, for example, \citet{Espinoza:2009ud} and \citet{Stead:2009oq}.
The uncertainties cited for the effective
wavelengths are based on best-estimates of the variablity of these
values for the case of Galactic center observations and the observing
conditions and instruments used in this work.

\begin{acknowledgements}
RS acknowledges the Ram\'on y Cajal programme of the Spanish
Ministerio de Ciencia e Innovaci\'on  (MICINN) and support by the
MICINN project AYA2007-64052. We thank the anonymous referee for his
helpful comments. We are grateful to Andrea Ghez and her group at UCLA
for providing detailed comments on this work.
\end{acknowledgements}

\bibliography{/Users/rainer_old/Documents/MyPapers/BibGC}

\appendix

\section{Effective wavelength \label{app:efflambda}}

We calculated the effective wavelengths of our
observations following equation (A3) of \citet{Tokunaga:2005jw}.  The
transmission curves for the NACO $H$ and $Ks$ filters were downloaded
from the instrument web
site\footnote{http://www.eso.org/sci/facilities/paranal/instruments/naco/index.html}.
Spectra of the atmospheric transmission were downloaded from the Gemini
telescope web
site\footnote{http://www.gemini.edu/node/10781?q=node/10789}. They are
based on models with the ATRAN software \citep{Lord1992}.  The
transmission curves for 3\,mm precipitable water vapor were used
because humidity was high during the $H$ and $Ks$ observations. 

We did not use any particular stellar atmosphere model for the
calculations. Instead, simple extincted blackbody models were used.
Table\,\ref{Tab:lambdaeff} lists the values of the effective
wavelength for various values of $A_{K}$ and effective temperatures of
the blackbodies. A power-law $A_{K}\propto\lambda^{-\alpha}$
\citep{Nishiyama:2009oj} was used to calculate the extinction at
different wavelengths. We assumed $\alpha_{H-Ks}=2.22$ for the
extinction between $H$ and $Ks$ , and $\alpha_{Ks-L'}=1.33$ for the
extinction between $Ks$ and $L'$ \footnote{This value of
  $\alpha_{Ks-L'}$ is from a preliminary estimate at the beginning of
  this work and is smaller than the value $\alpha_{Ks-L'}=1.38$ that
  we derived in the end. However, the associated change  of $\lambda_{\rm eff,
    L'}$ is minimal, see discussion of the uncertainties in the next paragraph.}.  The majority ($>90\%$) of stars in
the images analyzed in this work are of late-type, with a large
percentage of RC stars \citep[see, e.g.,][]{Buchholz:2009sp}. We
therefore adopt as effective wavelengths $\lambda_{\rm eff,
  H}=1.677\,\mu$m, $\lambda_{\rm eff, Ks}=2.168\,\mu$m, and
$\lambda_{\rm eff, L'}=3.636\,\mu$m for an average extinction of
$A_{KS}\approx2.5$.

The uncertainty introduced by the choice for the amount of
precipitable water vapor is low. Choosing $1.6$\,mm instead of $3.0$\,mm will
change the effective wavelengths by $\leq0.003\,\mu$m for $H$, $\leq0.001\,\mu$m
for $Ks$, and $\leq0.01\,\mu$m for $L'$.  Changing the power-law index for
the extinction curve between $H$ and $Ks$ to $\alpha_{H-Ks}=2.0$  \citep{Nishiyama:2009oj}
results in a change of $\leq0.003\,\mu$m for $\lambda_{\rm eff, Ks}$,
and $\leq0.009\,\mu$m for $\lambda_{\rm eff, H}$. Changing the
power-law index for the extinction curve between $Ks$ and $L'$ to
$\alpha_{Ks-L'}=1.37$ \citep{Nishiyama:2009oj} results in a change of $<0.001\,\mu$m for
$\lambda_{\rm eff, L'}$. The uncertainty in extinction introduces an
uncertainty of $\lesssim0.015\,\mu$m for $H$, $\lesssim0.01\,\mu$m for
$Ks$, and $\lesssim0.002\,\mu$m for $L'$. The
unknown effective temperature of the stars introduces uncertainties of
$\lesssim0.005\,\mu$m for $H$, $\lesssim0.006\,\mu$m for $Ks$, and
$\lesssim0.001\,\mu$m for $L'$.

We estimate the combined $1\,\sigma$ uncertainty of the effective
wavelengths to $0.018\,\mu$m for $H$, $0.012\,\mu$m for $Ks$, and
$0.012\,\mu$m for $L'$.

\begin{table}[htb]
\centering
\label{Tab:lambdaeff}
\caption{Effective wavelengths in units of $\mu$m for the NACO $H$,
  $Ks$, and $L'$ filters for blackbodies with different effective temperatures
and extinction, $A_{Ks}$.} 
\begin{tabular}{r|rrrr}
$H$ & & & & \\
\hline
              &       2.0$^{\mathrm{a}}$     &       2.5    &    3.0   & 3.5 \\
\hline
    3000$^{\mathrm{b}}$  &      1.674  &      1.681  &      1.688  &      1.695\\
    4700   &     1.669  &      1.677  &      1.684  &      1.691\\
  30000   &     1.664  &      1.672  &      1.680  &      1.687\\
\hline
$Ks$ & & & & \\
\hline
     3000  &      2.167   &     2.171  &      2.176   &     2.180\\
     4700  &      2.163   &     2.168  &      2.173   &     2.177\\
    30000 &       2.159  &      2.164 &       2.169  &      2.174\\
\hline
$L'$ & & & & \\
\hline
     3000  &       3.636   &    3.637   &  3.638    &  3.639   \\
     4700  &       3.635  &     3.636   &  3.637    &  3.638  \\
    30000 &       3.634  &     3.635   &  3.636    &  3.637    \\
\hline
\end{tabular}
\begin{list}{}{}
\item[$^{\mathrm{a}}$] Extinction in the $Ks$ band, $A_{Ks}$.
\item[$^{\mathrm{b}}$] Effective temperature.
\end{list}
\end{table}

\Online

\begin{table}
\begin{longtable}{lrrrrrrrrr}
  \caption{ \label{Tab:list} List of detected point sources in the
    NACO $H$, $Ks$, and $L'$-band observations.  We only include the
    first ten lines of the table in the printed edition of
    this work. }\\
  \hline \hline
  ID & R$_{\rm SgrA*}$$^{\mathrm{a}}$ & R.A.$^{\mathrm{b}}$ & Dec.$^{\mathrm{c}}$  & $H^{\mathrm{d}}$ &  $\Delta H^{\mathrm{e}}$  & $Ks^{\mathrm{f}}$ & $\Delta Ks^{\mathrm{g}}$  & $L'^{\mathrm{h}}$ &  $\Delta L'^{\mathrm{i}}$ \\
  & [$''$] & [$''$] & [$''$] &  &  & & & &  \\
  \hline
\endfirsthead
\caption{continued.}\\
\hline
\hline
  ID & R$_{\rm SgrA*}$$^{\mathrm{a}}$ & R.A.$^{\mathrm{b}}$ & Dec.$^{\mathrm{c}}$  & $H^{\mathrm{d}}$ &  $\Delta H^{\mathrm{e}}$  & $Ks^{\mathrm{f}}$ & $\Delta Ks^{\mathrm{g}}$  & mag$L'^{\mathrm{h}}$ &  $\Delta L'^{\mathrm{i}}$ \\
  & [$''$] & [$''$] & [$''$] &  &  & & & &  \\
\hline
\endhead
\hline
\endfoot
1 & 0.028 & 0.009 & -0.027 & 17.50 & 0.05 & 15.7 & 0.04 & 0.0$^{\mathrm{k}}$ & 0.0 \\ 
2 & 0.107 & 0.031 & -0.102 & 18.19 & 0.06 & 16.39 & 0.04 & 0 & 0 \\ 
3 & 0.163 & -0.006 & 0.163 & 16.00 & 0.02 & 14.13 & 0.01 & 12.84 & 0.03 \\ 
4 & 0.231 & 0.189 & 0.132 & 16.93 & 0.03 & 15 & 0.01 & 0 & 0 \\ 
5 & 0.245 & -0.231 & -0.081 & 18.69 & 0.07 & 16.75 & 0.05 & 0 & 0 \\ 
6 & 0.265 & 0.022 & -0.264 & 16.53 & 0.03 & 14.7 & 0.02 & 0 & 0 \\ 
7 & 0.321 & 0.296 & 0.124 & 16.47 & 0.02 & 14.61 & 0.01 & 0 & 0 \\ 
8 & 0.347 & -0.274 & 0.213 & 17.45 & 0.03 & 15.66 & 0.09 & 14.35 & 0.06 \\ 
9 & 0.354 & -0.053 & 0.35 & 17.22 & 0.03 & 15.39 & 0.02 & 0 & 0 \\ 
10 & 0.363 & 0.02 & -0.363 & 16.23 & 0.03 & 14.12 & 0.01 & 12.6 & 0.03 \\ 
\end{longtable}
\begin{list}{}{}
\item[$^{\mathrm{a}}$] Distance from Sgr\,A* projected on the sky.
\item[$^{\mathrm{b}}$] Offset in right ascension from
  Sgr\,A*. Uncertainties in the astrometric positions can be as large
  as $\sim0.1"$ because high precision astrometry was not at the focus
  of this work. The conversion to offsets from Sgr\,A* was done by
  assuming a pixel scale of $0.027"$ per pixel (ESO NACO manual). The
  camera rotation angle was assumed $0$\,deg east of north and no
  camera distortion solution was applied.
\item[$^{\mathrm{c}}$] Offset in declination from
  Sgr\,A*. For more information see also footnote $^{\mathrm{b}}$.
\item[$^{\mathrm{d}}$] Magnitude in the NACO $H$ band filter.
\item[$^{\mathrm{e}}$] Uncertainty in $H$, obtained by quadratically
  combining the formal uncertainty from PSF fitting, based on Gaussian
  read-out noise and Poisson photon noise, and the uncertainty due to
  the limited knowledge of the PSF. The $1\,\sigma$ uncertainty of the
  zero point, $\Delta ZP_{H}=0.06$, is not included in this column and
  has to be taken into account when considering \emph{absolute}
  photometry.
\item[$^{\mathrm{f}}$] Magnitude in the NACO $Ks$ band filter.
\item[$^{\mathrm{g}}$] Uncertainty in $Ks$. For more information see
  also footnote $^{\mathrm{e}}$. The $1\,\sigma$ uncertainty of the
  zero point is $\Delta ZP_{Ks}=0.06$.
\item[$^{\mathrm{h}}$] Magnitude in the NACO $L'$ band filter.
\item[$^{\mathrm{i}}$] Uncertainty in $L'$. For more information see
  also footnote $^{\mathrm{e}}$. The $1\,\sigma$ uncertainty of the
  zero point is $\Delta ZP_{L'}=0.15$.
\item[$^{\mathrm{k}}$] A value of $0.0$ in the table indicates that
  the corresponding measurement is not available.
\end{list}
\end{table}

\end{document}